\documentclass[aps,pre,floats,twocolumn,superscriptaddress,nobibnotes]{revtex4-2}
\usepackage[svgnames]{xcolor} 
\usepackage{graphicx,epsfig}
\usepackage{times} 
\usepackage{amssymb,amsmath,multirow,rotate,color}
\usepackage{xcolor}
\usepackage[normalem]{ulem}

\usepackage{todonotes}
\usepackage[colorlinks=true,
						linkcolor=blue,
						urlcolor=blue,
						citecolor=blue,
						bookmarks=true,
						pdfborder={0 0 0}]{hyperref}

\definecolor{albicocca}{rgb}{0.98, 0.7, 0.2}
\definecolor{internationalorange}{rgb}{1.0, 0.31, 0.0}
\definecolor{giocolor}{RGB}{0, 150, 100}

\usepackage{float}
\usepackage{lipsum} 
\usepackage{ragged2e}
\usepackage{soul}
\DeclareMathAlphabet\mathbfcal{OMS}{cmsy}{b}{n}
\newcommand{\av}[1]{ \langle #1 \rangle}

\newcommand{\braket}[1]{\langle #1\rangle}
\begin{document}

\title{Unveiling the impact of cross-order hyperdegree correlations in contagion processes on hypergraphs}

\author{Andr\'es Guzm\'an}
\affiliation{Network Science Institute, Northeastern University London, London E1W 1LP, United Kingdom}
\author{Federico Malizia}
\affiliation{Department of Network and Data Science, Central European University, Vienna, Austria}
\author{Istv\'an Z. Kiss}
\affiliation{Network Science Institute, Northeastern University London, London E1W 1LP, United Kingdom}
\affiliation{Department of Mathematics, Northeastern University, Boston, MA 02115, USA}
\date{\today}

\begin{abstract}

Contagion processes in social systems often involve interactions that go beyond pairwise contacts. Higher-order networks, represented as hypergraphs, have been widely used to model multi-body interactions, and their presence can drastically alter contagion dynamics compared to traditional network models. However, existing analytical approaches typically assume independence between pairwise and higher-order degrees, and thus study their roles in isolation. In this paper, we develop an effective hyperdegree model (EHDM) to describe Susceptible-Infected-Susceptible (SIS) dynamics on hypergraphs that explicitly captures correlations between the distribution of groups with different sizes. Our effective hyperdegree model shows excellent agreement with stochastic simulations across different types of higher-order networks, including those with heterogeneous degree distributions. We explore the critical role of cross-order  degree correlations, specifically, whether nodes that are hubs in pairwise interactions also serve as hubs in higher-order interactions. We show that positive correlation decreases the epidemic threshold and anti-correlation temporally desynchronizes infection pathways (pairwise and group interactions). Finally, we demonstrate that, depending on the level of correlation, the optimal control strategy shifts---from one that is purely pairwise- or higher-order-focused to one in which a mixed strategy becomes optimal.

\end{abstract}
\maketitle

\section{Introduction}

Complex networks have become a fundamental framework for modeling spreading processes and understanding a wide range of phenomena, including epidemics \cite{pastor2015epidemic,trapman2007analytical,kiss2017mathematics,pastor2002epidemic, barrat2008dynamical}, information diffusion \cite{daley1964epidemics, moreno2004dynamics, shi2025examining}, and the adoption of behaviours and social contagion \cite{hodas2014simple, centola2018experimental, watts2007influentials}. While this line of research has produced significant advances, recent years have seen growing interest in spreading phenomena that extend beyond pairwise interactions to encompass group interactions. This framework is especially important in the context of social dynamics, as reducing group interactions to a collection of pairwise connections can miss essential features of collective behavior. Empirical evidence shows that humans exhibit markedly different interaction patterns when engaging in groups compared to dyadic encounters \cite{greening2015higher, loersch2008influence}.

Accounting for multi-body interactions has been shown to profoundly affect dynamical processes such as contagion \cite{iacopini2019simplicial, de2020social, ferraz2024contagion}, synchronization \cite{bick2016chaos, skardal2019abrupt, millan2020explosive}, diffusion \cite{schaub2020random, carletti2020random, neuhauser2020multibody}, evolutionary dynamics \cite{civilini2024explosive,llabres2026emergence}, and social polarization \cite{perez2025social}. In spreading dynamics, the incorporation of higher-order interactions, i.e., interactions among groups of more than two individuals, reveals dynamical behaviors not captured by traditional pairwise network models. For reversible spreading processes such as the Susceptible–Infected–Susceptible (SIS) model, introducing three-body interactions, with links and triples representing different transmission pathways, leads to a variety of non-trivial phenomena. These include discontinuous phase transitions \cite{iacopini2019simplicial}, hysteresis loops \cite{gu2024epidemic}, explosive transitions \cite{matamalas2020abrupt, chen2025higher}, and bistability, where healthy and endemic states may coexist \cite{iacopini2019simplicial, ferraz2023multistability}.

More recently, considerable effort has been devoted to deepening our understanding of how structural characteristics influence what behaviors systems can exhibit.  In higher-order network representations, nodes can belong to hyperedges of different cardinality, leading to not one degree but a vector of hyperdegrees. The way in which hyperdegrees are distributed is fundamental in shaping dynamical processes unfolding on higher-order networks. For example, Landry et al. showed that the characteristic explosive transition previously reported can be suppressed by heterogeneity in pairwise interactions \cite{landry2020effect}. Conversely, heterogeneity in three-body interactions has been shown to play a key role in inducing explosive transitions  \cite{malizia2025disentangling}. 

The richness of these systems arises not only from the incremental inclusion of higher-order interactions, but also from the interplay between interactions of different orders. Beyond hyperdegree distributions, the arrangement of hyperedges has been shown to fundamentally shape epidemic dynamics~\cite{burgio2024triadic, kim2024higher, kim2023contagion, malizia2025hyperedge, malizia2025disentangling}. For instance, nested interactions, where lower-order interactions are embedded within higher-order ones~\cite{lamata2025hyperedge}, play a dual role in shaping epidemic onset: they can anticipate the emergence of epidemics while simultaneously suppressing abrupt transitions~\cite{malizia2026nested}. More broadly, nestedness influences not only macroscopic properties such as the final epidemic size, but also the dynamical pathways leading to explosive transitions~\cite{malizia2025disentangling}. Indeed, the mere presence of group interactions does not guarantee abrupt transitions in SIS processes~\cite{keating2025loops, malizia2025hyperedge}, as the structural arrangement of hyperedges, including correlations between different group sizes, profoundly shapes epidemic dynamics.

However, correlations between different orders of interaction can be examined from  multiple perspectives and are not limited to the presence of nested hyperedges. In many studies, the hyperdegree distributions, pairwise, three-body interactions, and higher-orders, are chosen independently from each order. This assumption is of course convenient from both algorithmic and analytical standpoints. Especially in the context of heterogeneous hyperdegree distributions, the correlations that arise when hyperdegrees are drawn from a joint distribution rather than treated as fully independent remain largely unexplored. Such correlations alter which nodes act as hubs at different interaction orders. Landry et al \cite{landry2020effect}. developed a hypergraph generative model alongside a degree-based spreading model and showed that positive correlations between the number of pairwise and group interactions a node participates in can promote the emergence of bistability and hysteresis \cite{landry2020effect}. Nonetheless, open questions remain regarding other types of correlations—for example, cases in which pairwise and higher-order hyperdegrees are anti-correlated, which may require more flexible modeling frameworks. Moreover, it is still unclear how different patterns of degree correlation shape the onset of the epidemic, influence the hierarchy of spread and how it impacts the optimality of control strategies based on hyperdegree of nodes.

In this paper, we overcome the limitations imposed by assuming independent hyperdegree distributions by introducing a configurational model with a joint distribution that naturally captures correlations between a node's participation in groups of different sizes. Furthermore, we develop an effective hyperdegree model (EHDM) for the Susceptible–Infected–Susceptible (SIS) process that accurately captures the dynamics on heterogeneous hypergraphs while accounting for correlations between interaction orders in the hyperdegree distribution. Through a systematic analysis of the epidemic threshold, quasi-equilibrium behavior, and time evolution, we provide a comprehensive characterization of how cross-degree correlations shape the system’s dynamical properties. Finally, we investigate how contagion propagates hierarchically through nodes belonging to different hyperdegree classes in a higher-order network with both pairwise and three-body interactions. Building on these insights, we show that optimal control strategies for targeting nodes based on their hyperdegree are dictated by the underlying correlations, with different correlation structures favoring purely pairwise, purely hyperdegree, or mixed targeting approaches.

\section{Configurational Higher-order network}\label{sec:conf-model}

A system exhibiting higher-order interactions can be represented by a hypergraph $H = (\mathcal{N}, \mathcal{E})$, where $\mathcal{N}$ is the set of $N = |\mathcal{N}|$ nodes and $\mathcal{E}$ is the set of $E = |\mathcal{E}|$ hyperedges representing their interactions.
Each hyperedge $e \in \mathcal{E}$ is a subset of $\mathcal{N}$ and can be characterized by its order $m = |e|-1$. For example, $m=1$ corresponds to pairwise interactions, $m=2$ corresponds to group interactions of 3 nodes, and so on. The generalized hyperdegree vector of a node $i$ is $\mathbf{k} = (k_1^i, k_2^i, \ldots, k_M^i)$, where $k_m^i$ denotes the number of $m$-hyperedges incident to a node~\cite{courtney2016generalized}. We define a joint probability distribution $\mathbf{P}(\mathbf{k})$ that gives the probability of a randomly chosen node having a hyperdegree vector $\mathbf{k} = (k_1, k_2, \ldots, k_M)$, where $M$ is the maximum order considered. This joint distribution has known marginal distributions $P_m(k_m)$ for each order of interactions $m$. 

A key aspect of our model is the choice of $\mathbf{P}(\mathbf{k})$ to describe correlations between orders of interaction. We define this relationship between orders as \textit{cross-order hyperdegree correlation}, quantifying how a node's participation in hyperedges of one order relates to its participation in hyperedges of another order. It is crucial to emphasize that this concept refers to the correlation between different hyperdegrees of a given node, rather than the correlation between a node's hyperdegrees and the hyperdegrees of its neighbors. The latter corresponds to degree assortativity which characterizes mixing patterns in the network but does not capture the relationship we investigate. Specifically, we examine correlations between hyperdegrees of different orders (e.g., how a node's 1-hyperdegree $k_1$ correlates with its 2-hyperdegree $k_2$), as opposed to correlations between hyperdegrees of the same order (e.g., whether nodes which are part of many 2-hyperedges tend to connect to other nodes that are also part of many 2-hyperedges).

For example, consider a higher-order network with only pairwise and three-body interactions, whose hyperdegree vector is $\mathbf{k} = [k_1, k_2]$. The cross-order hyperdegree correlation between the random variables $k_1$ and $k_2$ is then quantified by the Pearson correlation coefficient $\sigma \in [-1,1]$. For hypergraphs with more than two orders of interaction, this scalar parameter generalizes to a covariance matrix $\boldsymbol{\Sigma}$ that captures all pairwise correlations between hyperdegrees.

To construct a hypergraph, we need a hyperdegree vector for each node. We first draw $N$ hyperdegree vector samples from a multivariate normal distribution $\mathcal{N}_M(\mathbf{0}, \boldsymbol{\Sigma})$, where $\boldsymbol{\Sigma}$ encodes the desired cross-order correlations, yielding a vector $(z_1^n, z_2^n, \ldots, z_M^n)$ for each node $n$. Next, we apply the univariate standard normal cumulative distribution function (CDF) $\Phi(\cdot)$ component-wise to transform each element into uniform variables $(u_1^n, u_2^n, \ldots, u_M^n) = (\Phi(z_1^n), \Phi(z_2^n), \ldots, \Phi(z_M^n))$, where $u_m^n \in (0,1)$. Finally, we apply Inverse Transform Sampling~\cite{devroye2006nonuniform} to each component by computing $k_m^n = F_m^{-1}(u_m^n)$, where $F_m^{-1}$ is the inverse CDF of the desired marginal distribution $P_m(k_m)$. This method guarantees that the resulting hyperdegree vectors $\mathbf{k}^n = (k_1^n, k_2^n, \ldots, k_M^n)$ preserve the specified correlations while matching the desired marginal distributions.

Using the final hyperdegree vectors obtained, we employ a hypergraph configuration model and we generate hyperstubs—half-edges or unconnected elements—for each order $m$ according to the sampled hyperdegrees. For each interaction order $m$, we randomly group $m+1$ hyperstubs to create hyperedges of that order, repeating this process until all hyperstubs of order $m$ are exhausted. 

\begin{figure}[h!]
    \centering
    \includegraphics[width=0.9\linewidth]{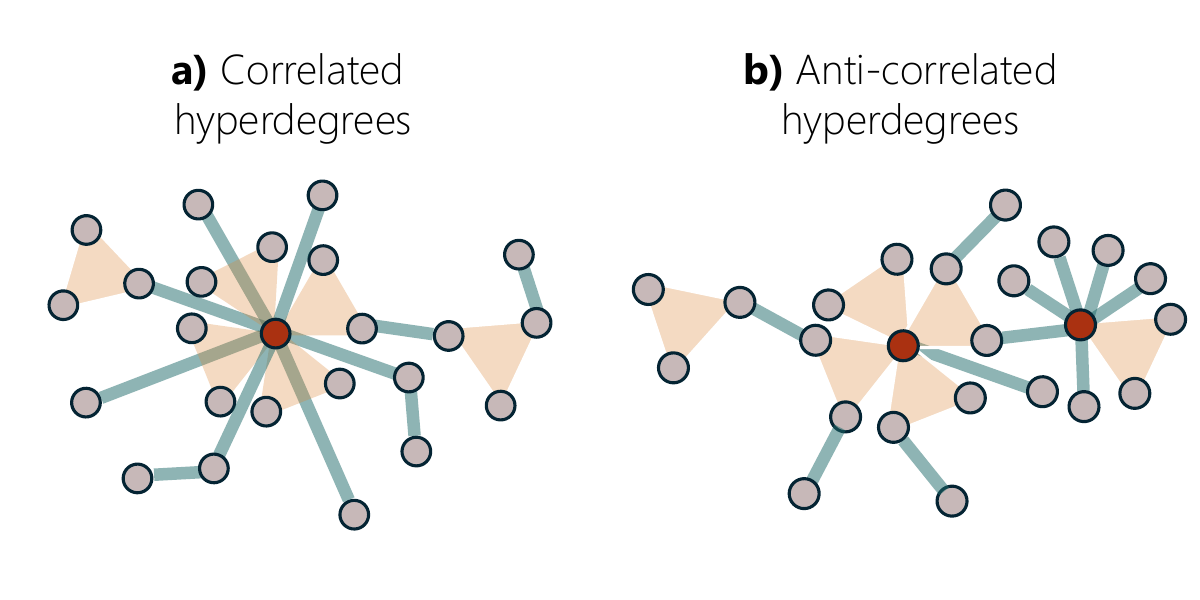}
    \caption{\textbf{Representation of hub distribution for correlated and anti-correlated systems:} This diagram depicts an illustrative representation of how cross-order hyperdegree correlations influence the distribution of highly connected nodes in a heterogeneous hypergraph. In (a) we show the positively correlated case, where the hub (red node) is highly connected in both group interactions and pairwise links. In (b) we show the negatively correlated case, where there are two distinct hubs (red nodes): one is highly connected in group interactions with very few pairwise links while the other is highly connected in pairwise links with few group interactions.}
\label{fig:diagram}
\end{figure}

With this formulation, we can generate hypernetworks with arbitrary marginal degree distributions while incorporating tunable cross-order hyperdegree correlations. Within this framework, for hypergraphs with heterogeneous hyperdegree distributions, the organization of highly connected nodes (hubs) depends on the nature of the cross-degree correlations. For example, in a highly correlated system, the same nodes act as hubs across all interaction orders, whereas in an anti-correlated system, each interaction order is dominated by its own distinct set of hubs. An illustrative example of how the role of hubs changes is shown in Fig. \ref{fig:diagram} for a system with pairwise and three-body interactions. With this flexible configurational hyperdegree model we are able to study how hyperdegree correlation may influence the emergent behavior in spreading processes. Further details on the configurational model are provided in the Supplementary Material (SM) \cite{supplementary}.

\section{SIS Higher-order effective hyperdegree model}\label{sec:SIS_effdegree}

Having introduced a configurational model that takes into account cross-order hyperdegree correlations, we are now interested in studying their effects in social contagion/epidemic process. For this, we consider an SIS process on hypergraphs, where nodes can be either susceptible ($S$) or infected ($I$). An infection occurs when a susceptible node belongs to an $m$-hyperedge whose other $m-1$ nodes are infected. Each interaction order is associated with a characteristic infection rate $\beta_m$, while the recovery rate $\gamma$ is assumed to be independent of hypernetwork structure. In what follows, we consider two orders of interaction: $ m = 1$, representing pairwise (two-node) interactions, and $m = 2$, representing three-body interactions. However, our framework can be straightforwardly extended to higher-order interactions. 

The state of each hyperedge depends on the states of its constituent nodes. For instance, a three-node hyperedge can have configurations ranging from all nodes susceptible to all nodes infected. To capture the dynamics accurately, we must track not only individual node states but also the distribution of hyperedge states, accounting for dynamical correlations that emerge during the epidemic process.
The model we introduce here is a generalization of the effective-degree framework previously developed for network-based epidemics in~\cite{lindquist2011effective}. In the original effective degree model, designed for epidemics with only pairwise interactions, variables such as $S_{s,i}$ track the number of susceptible nodes with $s$ links to other susceptible nodes and $i$ links to infected nodes. We extend this framework to higher-order interactions by considering all possible state configurations of the different types of hyperedges incident to a node. A similar approach was previously introduced in ~\cite{chen2023composite} in a discrete-time framework.

Our model tracks nodes according to both their infection state (susceptible or infected) and the complete state of their neighborhood. For each node, we account for: (i) the number of pairwise links to susceptible and infected nodes, and (ii) the number of three-body hyperedges containing different combinations of susceptible and infected neighbors (specifically, hyperedges with 2$S$, 1$S$-1$I$, or 2$I$ configurations).

As a result, a randomly selected node $u$ is characterized by its neighborhood, that is, the state of nodes connected to $u$ is described by a \emph{neighborhood vector}, $\mathbf{n}_u = [s,i,x,y,z]$ where: $s$ - number of links to susceptible nodes, $i$ -  number of links to infected nodes, $x$ - number of three-body interactions in which the other two nodes are susceptible, $y$ - number of groups in which one node is susceptible and the other infected, and $z$ - number of groups in which both other nodes are infected.  We describe the system’s dynamics by grouping together nodes that share the same neighborhood vector. Let $S_{s,i}^{x,y,z}(t)$ denote the total number of susceptible nodes with neighborhood vector $\mathbf{n}_u = [s,i,x,y,z]$ at time $t$, and $I_{s,i}^{x,y,z}(t)$ stands for the number of infected nodes with the same neighborhood configuration.  

The model requires additional variables beyond individual node tracking: we introduce counts for pairwise interactions between nodes in states A and B (denoted $[AB]$), as well as counts for three-body hyperedges with nodes in states A, B, and C (denoted $[ABC]$). These hyperedge-level quantities group interactions by their state composition rather than by the individual nodes they contain. For example, $[SS]$ counts pairs of two susceptible nodes, and $[SII]$ counts three-body interactions with one susceptible and two infected nodes. 

Finally, to capture the full dynamics, we must account for how neighbors of a focal node can become infected through their other hyperedge connections. When a susceptible neighbor of a focal node participates in additional hyperedges beyond the one containing the focal node, it can become infected through these. This requires tracking configurations where two hyperedges share a common node—one hyperedge containing our focal node, and another providing an additional route of infection. For instance, consider the structure $I\underline{S}II$, where the underlined susceptible node is shared between two hyperedges: it connects to an infected node through one hyperedge while simultaneously participating in a three-body hyperedge with two other infected nodes. This shared node can become infected through either hyperedge, and we need these overlapping structure counts to correctly compute the rate at which the neighborhood of any focal node changes due to  infection.

The evolution of $S_{s,i}^{x,y,z}(t)$ and $I_{s,i}^{x,y,z}(t)$ depends on both the state of the focal node and the states of its neighbors. For example, if a susceptible node $u$ has neighborhood vector $\mathbf{n}_u = [s,i,x,y,z]$, the state of $u$ can change if: $u$ becomes infected by one of its infected neighbors; $u$’s neighborhood vector changes if one of its susceptible neighbors becomes infected or one of its infected neighbors recovers. To connect neighborhood dynamics with hyperedge counts, we define a joint variable that tracks both node state and local hyperedge arrangement structure. Specifically, $[A\underline{B}CD_{s,i}^{x,y,z}]$ 
denotes the number of nodes in state $D$ with neighborhood vector $\mathbf{n}_u = [s,i,x,y,z]$ that belong to a 2-hyperedge with two other nodes in states $C$ and $B$, where the shared node $B$ also participates in a 1-hyperedge with a node in state $A$.  For example, $[I\underline{S}SS_{s,i}^{x,y,z}]$ represents the number of susceptible nodes with neighborhood state $\mathbf{n}_u = [s,i,x,y,z]$ that belong to a three-body interactions with two other susceptible nodes, one of which is also connected by a pairwise link to an infected node.

The central assumption of the effective hyperdegree model is that all susceptible neighbors of a node are statistically equivalent. Specifically, given a focal node, we assume that the rate at which any of its susceptible neighbors,  connected via a hyperedge, becomes infected is identical to the rate of any other susceptible neighbor in an equivalent hyperedge state configuration.  Under this assumption, number of configurations such as $[I\underline{S}SS_{s,i}^{x,y,z}]$ can be approximated using global hyperedge arrangement counts. In particular,

\begin{equation}
    [I\underline{S}SS_{s,i}^{x,y,z}] \;\approx\; 
\frac{[I\underline{S}SS]}{[SSS]} \, xS_{s,i}^{x,y,z},
\end{equation}
where, the term $xS_{s,i}^{x,y,z}$ gives the number of three-body interactions where each node is susceptible, while $\frac{[I\underline{S}SS]}{[SSS]}$ gives the probability that a susceptible node already in a 3-body interaction with two other susceptible is also part of a pairwise interaction with another infected node. The same logic applies to infected nodes and to any other combination of hyperedges in any other state. The total counts of individual hyperedges and hyperedge arrangements can be expressed in terms of the node-level quantities $S_{s,i}^{x,y,z}$ and $I_{s,i}^{x,y,z}$. For example, the total number of susceptible pairs is $[SS] = \sum_{\mathbf{n}_u} s \, S_{s,i}^{x,y,z}$, where the sum runs over all possible neighborhood vectors $\mathbf{n}_u$. Similarly, the count of the two-hyperedge arrangements $[I\underline{S}SS]$ is $[I\underline{S}SS] = \sum_{\mathbf{n}_u} i \, x \, S_{s,i}^{x,y,z}$. With these definitions in place, we can now write the evolution equations governing the dynamics of $S_{s,i}^{x,y,z}$ and $I_{s,i}^{x,y,z}$.

\begin{widetext}
\begin{equation}
    \begin{split}
        \dot{S}_{s,i}^{x,y, z} =& -(\beta_1i+\beta_2z) S_{s,i}^{x,y,z} + \gamma I_{s,i}^{x,y,z}
         + \Big(\beta_1 \frac{[I\underline{S}S]}{[SS]}+\beta_2\frac{[II\underline{S}S]}{[SS]}\Big)\Big( (s+1)S_{s+1,i-1}^{x,y,z} -s S_{s,i}^{x,y,z} \Big)  
         \\& + \Big(\beta_1\frac{[I\underline{S}SS]}{[SSS]}+\beta_2\frac{[II\underline{S}SS]}{[SSS]}\Big)\Big( (x+1)S_{s,i}^{x+1,y-1,z} -x S_{s,i}^{x,y,z} \Big)
        \\&+ \Big(\beta_1\frac{[I\underline{S}IS]}{ISS}+\beta_2\frac{[II\underline{S}IS]}{ISS}\Big)\Big( (y+1)S_{s,i}^{x,y+1,z-1} -y S_{s,i}^{x,y,z} \Big)\\ & + \gamma \Big(-(i+y+2z)S_{s,i}^{x,y,z} + (i+1)S_{s-1,i+1}^{x,y,z} +  (y+1)S_{s,i}^{x-1,y+1,z} + 2(z+1)S_{s,i}^{x,y-1,z+1} \Big) \\
        \dot{I}_{s,i}^{x,y, z} =  &(\beta_1i+\beta_2z) S_{s,i}^{x,y,z} - \gamma I_{s,i}^{x,y,z} 
        + \Big(\beta_1+ \beta_1\frac{[I\underline{S}I}{IS}+\beta_2\frac{[II\underline{S}I]}{IS}\Big)\Big( (s+1)I_{s+1,i-1}^{x,y,z} -s I_{s,i}^{x,y,z} \Big) 
        \\ & + \Big(\beta_1\frac{[I\underline{S}IS]}{[ISS]}+\beta_2\frac{[II\underline{S}IS]}{[ISS]}\Big)\Big( (x+1)I_{s,i}^{x+1,y-1,z} -x I_{s,i}^{x,y,z} \Big) 
        \\&+ \Big(\beta_2+ \beta_1\frac{[I\underline{S}II}{IIS}+\beta_2\frac{[II\underline{S}II}{IIS}\Big)\Big( (y+1)I_{s,i}^{x,y+1,z-1} -y I_{s,i}^{x,y,z} \Big) + 
        \\ & + \gamma \Big(-(i+y+2z)I_{s,i}^{x,y,z} + (i+1)I_{s-1,i+1}^{x,y,z} + (y+1)I_{s,i}^{x,y+1,z-1} + 2(z+1)I_{s,i}^{x,y-1,z+1} \Big)
    \end{split}
    \label{Eff_degree}
\end{equation}
\end{widetext}

We stress that the novelty of our model in equation~\eqref{Eff_degree} is the ability to track dynamical correlations and hyperdegree heterogeneity. Additionally, cross-order correlations between the pairwise and three-body  distributions around a node can be incorporated directly through the initial conditions, i.e. setting $S_{s,0}^{x,0,0}(t=0)$ to desired values. 

\section{SIS Higher-order compact effective hyperdegree model}\label{sec:SIS_compeffdegree}

The ability of the model to capture heterogeneous and cross-order correlated hypergraphs comes at a cost. In particular, the model is inherently high-dimensional, as also noted in~\cite{chen2023composite}. The number of variables grows rapidly with the range of hyperdegrees, since all configurations $S_{s,i}^{x,y,z}$ and $I_{s,i}^{x,y,z}$ must be tracked explicitly. To reduce the complexity of the full system, we develop a compact effective hyperdegree approximation~\cite{kiss2017mathematics}.

The key assumption underlying this simplification is that the states of hyperedges of a central node are independently distributed. This is, for a node participating in $k_1$ pairwise interactions, each neighbor is independently susceptible or infected with equal probability. Similarly, for $k_2$ three-body interactions, each hyperedge is assumed to independently occupy any of its possible configurations: two susceptible nodes, one susceptible and one infected node, or two infected nodes.

Under this approximation, nodes are characterized just by their number of 1-hyperedges and 2-hyperedges, hence we can define a new set of variables $S_{k_1}^{k_2}$ and $I_{k_1}^{k_2}$ for susceptible and infected nodes with $k_1$ and $k_2$ hyperdegrees, respectively. The full set of effective hyperdegree variables $S_{s,i}^{x,y,z}$ and $I_{s,i}^{x,y,z}$ can then be approximated by multinomial distributions over these compact variables:
\begin{equation}
S_{s,i}^{x,y,z} \approx \bigg[ \frac{k_1!}{s!\, i!} \braket{I}^i \braket{S}^s \bigg]
\bigg[ \frac{k_2!}{x!\, y!\, z!} \braket{X}^x \braket{Y}^y \braket{Z}^z \bigg] S_{k_1}^{k_2},
\end{equation}

where $\braket{S}, \braket{I}, \braket{X}, \braket{Y}, \braket{Z}$ denote probabilities determined by the global counts of hyperedges in each state:
\begin{equation}
    \begin{split}
        &\braket{S} = \frac{[SS]}{\sum k_1 S_{k_1}^{k_2}}, \hspace{0.3cm}  
        \braket{I} = \frac{[SI]}{\sum k_1 S_{k_1}^{k_2}}, \\
        &\braket{X} = \frac{[SSS]}{\sum k_2 S_{k_1}^{k_2}}, \hspace{0.3cm}
        \braket{Y} = \frac{[SSI]}{\sum k_2 S_{k_1}^{k_2}}, \\
        &\braket{Z} = \frac{[SII]}{\sum k_2 S_{k_1}^{k_2}}.
    \end{split}
\end{equation}
Here, $\sum = \sum_{k_1,k_2}$ denotes summation over all hyperdegree classes.

The time evolution of the compact variables is obtained by summing over the equations of the original model, for example $\dot{S}_{k_1}^{k_2} = \sum_{s,i} \sum_{x,y,z} \dot{S}_{s,i}^{x,y,z}$ with $s+i =k_1$ and $x+y+z=k_2$.
The full reduction of the hyperdegree model to the compact version can be found in the SM \cite{supplementary}. The system for the newly proposed variables $S_{k_1}^{k_2}(t)$ and $I_{k_1}^{k_2}(t)$ can be closed and made self-consistent in terms of the compact variables and the hyperedge states by treating the global probabilities as functions of the hyperedge state counts. The final compact effective hyperdegree dynamics are then given by:

\begin{equation}
    \begin{split}
    \dot{S}_{k_1}^{k_2} &= - \bigg(\beta_1k_1\braket{I} + \beta_2 k_2\braket{Z}\bigg) S_{k_1}^{k_2} + \gamma \big( N_{k_1}^{k_2} - S_{k_1}^{k_2}\big)\\
        \dot{[SI]}& = \gamma[II]-(\gamma +\beta_1) [SI] 
        \\&+ \bigg( 1 - 2\braket{I}\bigg) \bigg(\beta_1\braket{I}D + \beta_2\braket{Z}C\bigg)\\
        \dot{[II]} &= -2\gamma[II] + \beta_1[SI] + \braket{I} \bigg(\beta_1\braket{I}D + \beta_2\braket{Z}C\bigg)\\
        \dot{[SSI]} &= \gamma \bigg( [SII]-[SSI] \bigg) 
        \\&+ \bigg( 1 - \braket{Z} -2\braket{Y}\bigg) \bigg(\beta_1\braket{I}C + \beta_2\braket{Z}E\bigg) \\
        \dot{[SII]}& = \gamma[III] -(2\gamma +\beta_2) [SII] 
        \\& + \bigg( \braket{Y} - \braket{Z}\bigg) \bigg(\beta_1\braket{I}C + \beta_2\braket{Z}E\bigg)\\
        \dot{[III]}& = -3\gamma[III] + \beta_2 [SII] 
        \\&+ \braket{Z} \bigg(\beta_1\braket{I}C + \beta_2\braket{Z}E\bigg)\\
         \braket{I} &= \frac{[SI]}{A}, \hspace{0.5cm}  \braket{Z}= \frac{[SII]}{B}, \hspace{0.5cm}  \braket{Y}= \frac{[SSI]}{B}
    \end{split}
\end{equation}

with
\begin{equation}
    \begin{split}
        &A = \sum k_1 S_{k_1}^{k_2}, \quad B = \sum_{k_1,k_2} k_2 S_{k_1}^{k_2}, \\
        &C = \sum k_1 k_2 S_{k_1}^{k_2}, \quad D = \sum k_1 (k_1 - 1)S_{k_1}^{k_2}, \quad \\
        & E = \sum k_2 (k_2-1) S_{k_1}^{k_2}.
    \end{split}
\end{equation}

This compact formulation preserves the essential features of higher-order interactions while dramatically reducing the number of equations, making the system tractable for analysis. Specifically, if $K_1$ and $K_2$ are the maximum 1-hyperdegree and 2-hyperdegree respectively, the original model requires a number of equations that scales as $\mathcal{O}(K_1^2 K_2^3)$, while the compact version scales as $\mathcal{O}(K_1 K_2)$, details of this reduction are presented in SM \cite{supplementary}. For example, with $K_1 = 20$ and $K_2 = 10$, the reduction factor is approximately 220, meaning the full model requires tracking 220 times more state variables than the compact model. This complexity reduction is particularly important for investigating how cross-order hyperdegree correlations shape contagion processes in heterogeneous hypergraphs, as the model remains computationally feasible even for networks with large hyperdegree ranges. From this point onward, we refer to this compact version as the Effective Hyperdegree Model (EHDM).

\section{Results}\label{sec:results}

Having defined our mathematical framework, we first validate our model by testing on three types of hypernetworks: (i) a regular hypernetwork, in which all nodes have the same number of pairwise interactions $k_1$ and the same number of three-body interactions $k_2$; (ii) a hypernetwork with Poisson hyperdegree distribution for both links and groups, with average hyperdegrees $\langle k_1 \rangle$ and $\langle k_2 \rangle$, respectively; (iii) a hypernetwork with a truncated power-law distribution for both pairs and groups with characteristic exponent $\nu_1$ for $P(k_1)$ and $\nu_2$ for $P(k_2)$. For this initial case, we consider the distributions to be independent. As a consequence, the hyperdegree distributions are uncorrelated. All hypernetworks were generated using the higher-order configuration model described in Section~\ref{sec:conf-model}.

\begin{figure}[ht!]
    \centering
    \includegraphics[width=0.95\linewidth]{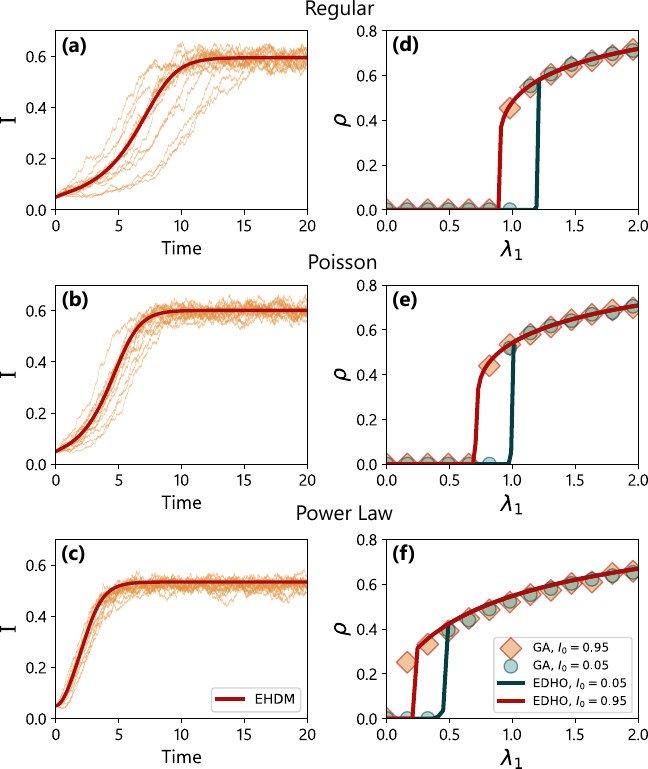}
    \caption{\textbf{EHDM vs simulation comparison:} We compare the effective hyperdegree model against stochastic simulations across three distinct hypergraph classes: a regular hypergraph where all nodes have $k_1=5$ and $k_2=3$; a hypergraph with Poisson-distributed hyperdegrees with means $\langle k_1 \rangle=5$ and $\langle k_2 \rangle=3$; and a hypergraph exhibiting truncated power-law hyperdegree distributions with characteristic exponents $\nu_1=2.5$ for pairwise interactions and $\nu_2=2.25$ for three-body interactions. All simulations use $N=1000$ nodes and a recovery rate $\gamma=1$. Panels (a--c) compare the temporal evolution of the infected population fraction between the effective hyperdegree model (red line) and Gillespie simulations (orange lines), with infection rates $\lambda_1=1.5$ and $\lambda_2=3$. Panels (d--f) show phase diagrams of the final infected fraction $\rho$ as a function of $\lambda_1$, comparing stochastic simulations (diamonds and circles obtained through 100 realizations of the process, confidence intervals not shown as often these are covered by the markers) with the EHDM (solid lines). In these cases we set $\lambda_2=2.5$, and  different colors indicate different initial numbers of infected nodes, and indicated in the legend. }
\label{fig:results1}
\end{figure}

Gillespie simulations were performed on these higher-order hypernetworks to compare with our model predictions. To facilitate the analysis, we define the effective infection rates $\lambda_1 = \langle k_1 \rangle \beta_1/\gamma$ and $\lambda_2 = \langle k_2 \rangle \beta_2/\gamma$. Figure~\ref{fig:results1} shows the comparison between the compact effective hyperdegree model and simulations for both the temporal evolution of the proportion of infected nodes and the phase diagram of prevalence (i.e., the final proportion of infected individuals) as a function of $\lambda_1$.

In Figure~\ref{fig:results1}(a--c), we observe that the EHDM demonstrates excellent agreement with simulations in describing the temporal evolution of the proportion of infected individuals across all hypergraphs. Additionally, in Figure~\ref{fig:results1}(d--f) we show the prevalence as a function of $\lambda_1$ for two distinct initial conditions: $I_0 = 0.05$ and $I_0 = 0.95$. We observe the typical behavior of SIS processes in the presence of higher-order interactions~\cite{iacopini2019simplicial,battiston2021physics}, namely, the presence of three regions in the phase diagram: (i) an absorbing region, where no outbreak occurs regardless of initial conditions; (ii) a bistable region, where two stable equilibria coexist and the system converges to one depending on the initial conditions, specifically, for low $I_0$ the epidemic dies out, while for sufficiently high $I_0$ a finite proportion of the population remains infected in the steady state; and (iii) an active region or endemic phase, where a finite proportion of the system is infected in the steady state and the system converges to the same stable equilibrium regardless of initial conditions. These regions are separated by two transitions: the backward transition, corresponding to the transition from the absorbing to the bistable regime, and the forward transition, corresponding to the transition from the bistable to the active regime.

Despite all structures having similar average hyperdegree values, we observe that the heterogeneous structures exhibit earlier onset of both the backward and forward transitions, consistent with previous findings in the literature\cite{landry2020effect, malizia2025disentangling}.

\subsection{Effect of hyperdegree heterogeneity} \label{sec:results_heterog}

As demonstrated in the previous section, the compact effective hyperdegree model exhibits excellent performance in capturing both the temporal evolution and steady-state behavior across higher-order networks with diverse hyperdegree distribution characteristics. Leveraging this accuracy, we now examine how hyperdegree heterogeneity influences epidemic outcomes. To isolate the effect of hyperdegree heterogeneity, we consider uncorrelated hyperdegree distributions generated using the configuration higher-order network model described in Section~\ref{sec:conf-model}. From this section onward, we focus on higher-order networks with negative binomial hyperdegree distributions for both pairwise and three-body interactions. This choice provides flexibility, as adjusting the distribution parameters $(r,p)$ directly controls the variance and thus the heterogeneity in connectivity. A detailed description of how we use this distribution to generate hypergraphs with varying heterogeneity is provided in the SM \cite{supplementary}.

\begin{figure}[h!]
    \centering
    \includegraphics[width=1\linewidth]{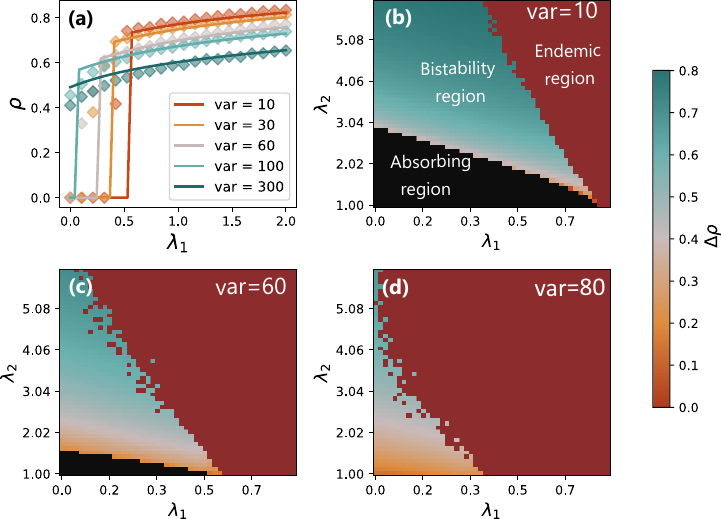}
    \caption{\textbf{Role of hyperdegree heterogeneity:} Panel (a) shows the final epidemic size as a function of the effective infection rate $\lambda_1$ for five different hypernetworks, comparing the average epidemic size from 100 stochastic simulations   (diamonds) with the effective hyperdegree model (solid lines). All networks have $N = 1000$ nodes with negative binomial hyperdegree distributions for both pairs and triples, average hyperdegrees $\langle k_1 \rangle \approx 5$ and $\langle k_2 \rangle \approx 3$, and varying hyperdegree distribution variance (10, 30, 60, 100, and 300). Panels (b--d) display heat-maps of the prevalence difference $\Delta\rho$ between two epidemic processes with identical infection and recovery rates but different initial numbers of infected individuals, for networks with variance 10, 60, and 80. The recovery rate is set to $\gamma = 1$ for all cases.}
\label{fig:results2}
\end{figure}

In Figure~\ref{fig:results2}(a), we show the phase diagram for five different higher-order networks with increasing variance (set to be equal for both pairwise and three-body interactions within each hypernetwork). For all cases, we consider a small number of initial infected nodes $I_0 = 0.05$. We present both stochastic simulations and the solution of our model to further support its validity. From the results in this panel, we observe that increasing variance decreases the epidemic threshold associated with the forward transition of the system. This result is consistent with classical findings for pairwise networks, where heterogeneous degree distributions are known to lower the epidemic threshold ~\cite{pastor2001epidemic}.

Furthermore, we investigate the impact of heterogeneity on the emergence of bistability. To this end, we define $\Delta\rho$ as the difference in the stationary states of infected densities between two trajectories: $\rho_1$, obtained from scenarios initialized with a large fraction of infected nodes ($I_0 = 0.95$) corresponding to the backward branch, and $\rho_2$, obtained from simulations initialized with a small fraction of infected nodes ($I_0 = 0.05$) corresponding to the forward branch. This is equivalent to taking the difference between the curves shown in Figure~\ref{fig:results1}(d--f). In Figure~\ref{fig:results2}(b--d), we show the values of $\Delta\rho$ (color-coded) in the ($\lambda_1,\lambda_2$) space calculated through the effective hyperdegree model, each panel with increasing variance in hyperdegree distribution. We can observe three distinct regions: the black area where $\Delta\rho = 0$ (absorbing phase); the red area where $\Delta\rho = 0$ and both branches lead to similar outbreak sizes (endemic phase); and the colored area where $\Delta\rho > 0$, indicating bistability. From these panels, we observe that higher heterogeneity shrinks the region of bistability. Our findings are consistent with those reported in ~\cite{landry2020effect}.

\subsection{Effect of cross-order hyperdegree correlations} \label{sec:results_cross_corr}

As shown in the previous section, heterogeneous  hyperdegree distributions can strongly influence epidemic outcomes due to the presence of highly connected nodes in hypergraphs. In this section, we investigate how correlations between different orders of interaction modify the role of these nodes and their impact on spreading dynamics. To this end, we employ the configuration model to construct hypergraphs with a joint negative binomial degree distribution, where cross-order correlations are controlled by covariance $\sigma$. Specifically, $\sigma = 1$ corresponds to a fully correlated system in which nodes with large pairwise hyperdegree also belong to many three-body interactions; $\sigma = 0$ represents an uncorrelated case with independent  hyperdegrees; and $\sigma = -1$ denotes an anti-correlated system where nodes with high pairwise hyperdegree tend to have small number of three-body interactions, and vice versa. 

\begin{figure*}[ht!]
    \centering
    \includegraphics[width=0.9\linewidth]{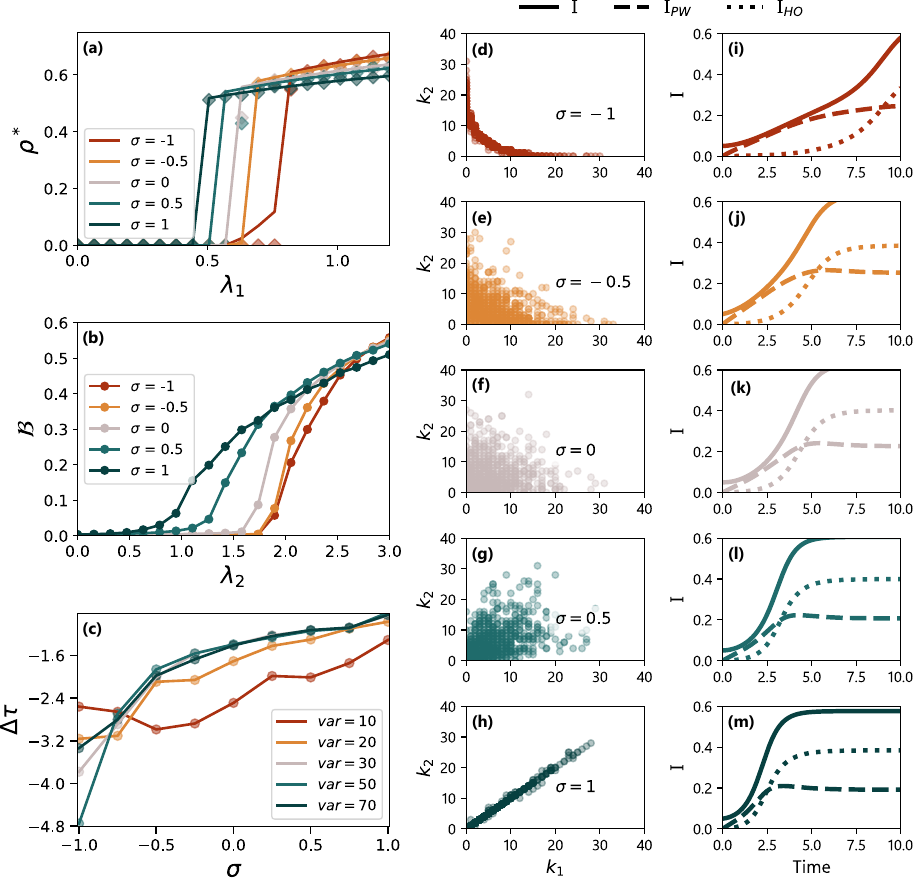}
    \caption{\textbf{Effect of cross-order hyperdegree correlation:} Panel (a) shows the final epidemic size as a function of infection rate $\lambda_1$ for heterogeneous hypernetworks with varying cross-order hyperdegree correlations ($\sigma = -1$ to $1$, shown in different colors), where the average of 100 stochastic simulations (diamonds) are compared with the effective hyperdegree model (solid lines) for $\lambda_2 = 3$. Panel (b) presents the bistability index $\mathcal{B}$ as a function of $\lambda_2$ across the same range of correlation values for $\lambda_1 = 0.9$. Panel (c) illustrates the difference between temporal centroids of pairwise infections ($I_{\text{PW}}$) and group infections ($I_{\text{HO}}$) as a function of $\sigma$ for five networks with different heterogeneity levels, quantified by the hyperdegree distribution variance indicated in the legend. Panels (d--h) display the hyperdegree distributions for networks with $\sigma = -1, -0.5, 0, 0.5$, and $1$, respectively. Panels (i--m) show the temporal evolution of the total infected fraction $I$, along with the contributions from pairwise ($I_{\text{PW}}$) and group ($I_{\text{HO}}$) infections, for hypernetworks with cross-order correlations corresponding to the hyperdegree distributions shown in panels (d--h). All hypernetworks contain $N = 1000$ nodes with negative binomial hyperdegree distributions for both pairwise and triple interactions, average hyperdegrees $\langle k_1 \rangle \approx \langle k_2 \rangle \approx 6$, and variance $\approx 30$ for both interaction orders, except in panel (c) where variance varies as indicated in the legend. Disease parameters for panels (c) and (i--m) are $\lambda_1 = 0.9$, $\lambda_2 = 3$, and $\gamma = 1$ for all cases.}
\label{fig:results3}
\end{figure*}

We consider five hypergraphs that share the same mean and variance of the  hyperdegree distribution (and thus identical $r$ and $p$ parameters of the negative binomial distribution) but differ in the covariance $\sigma$, ranging from $-1$ to $1$. In Fig.~\ref{fig:results3}(d--h), we show examples illustrating how $\sigma$ shapes the joint  hyperdegree distribution for $\sigma=-1,-0.5,0,0.5,1$. First, we fixed $\lambda_2 = 3$ and ran Gillespie simulations together with the compact effective hyperdegree model for range of $\lambda_1$ values.
Figure~\ref{fig:results3}(a) displays the final epidemic size as a function of $\lambda_1$ for the five hypergraphs previously mentioned. We find that the agreement of the compact effective hyperdegree model with stochastic simulations is preserved independently of the cross-order hyperdegree correlation.

A noticeable effect of cross-order hyperdegree correlation shown in Fig.~\ref{fig:results3}(a) is how it changes the epidemic threshold of the forward transition. Specifically, the critical value of $\lambda_1$ at which the system transitions to the endemic phase decreases steadily as the cross-order hyperdegree correlation increases, with the totally anti-correlated case ($\sigma=-1$) producing the highest threshold. This implies that systems in which highly connected nodes differ across interaction orders require higher infectivity for an epidemic to emerge. Furthermore, above the epidemic threshold, the anti-correlated case achieves the highest prevalence among all cases. This is likely because in the anti-correlated case, the system contains more hubs overall, connectivity is distributed across distinct node sets, increasing the effective coverage of transmission pathways. In contrast, in the correlated case, the same nodes serve as hubs for both interaction orders, reducing the total number of distinct hubs in the system.

Furthermore, we investigate how cross-order hyperdegree correlations influence the bistability described in the previous sections. Using the compact effective hyperdegree model, we computed the final epidemic size over a range of values of $\lambda_1$ and $\lambda_2$ for two distinct initial conditions: one with a small number of initially infected nodes and another with a large number. We rely exclusively on the compact effective hyperdegree model, as we have previously demonstrated its excellent agreement with stochastic simulations. For each value of $\lambda_2$, we calculated the bistability index, defined as the maximum difference between the final epidemic sizes obtained from the two initial conditions over all values of $\lambda_1$, following the definition introduced by Landry et al.~\cite{landry2020effect}. Figure~\ref{fig:results3}(b) shows the bistability index ($\mathcal{B}$) as a function of $\lambda_2$ for the hypergraphs previously considered. Across all levels of cross-order hyperdegree correlation, increasing $\lambda_2$ induces a transition from a regime with $\mathcal{B}=0$, where bistability is absent, to a regime in which the bistability index becomes nonzero. This behavior is consistent with previous findings~\cite{lamata2025hyperedge}. Importantly, stronger hyperdegree correlations reduce the critical value of $\lambda_2$ required to induce bistability, with the fully correlated case ($\sigma=1$) exhibiting the lowest threshold.

To further investigate the effect of cross-order hyperdegree correlation, we examine the temporal evolution of the system. 

For a more detailed description of the process, it is useful to separate the contributions from each order of interaction to the global proportion of infected nodes. Following \cite{malizia2025disentangling}, we define $I_{PW}$ as the contribution from pairwise interactions and $I_{HO}$ as the contribution from three-body interactions, such that:

\begin{equation}
\begin{split}
    &\dot{I}_{\rm PW} = \sum_{k_1,k_2} \bigg( \beta_1 k_1 \av{I}S_{k_1}^{k_2} \bigg)  - \gamma I_{\rm PW}\\
    &\dot{I}_{\rm HO} = \sum_{k_1,k_2} \bigg( \beta_2 k_2 \av{Z}S_{k_1}^{k_2} \bigg)  - \gamma I_{\rm HO} 
\end{split}
\end{equation}

In Fig.~\ref{fig:results3}(i--m), we show the temporal evolution of the epidemic process for $\lambda_1 = 1.1$~and $\lambda_2 = 2$~under different levels of cross-order hyperdegree correlation (corresponding to the cases shown in Fig.~\ref{fig:results3}(d--h)). We observe that in the anti-correlated cases, the spread driven by three-body interactions ($I_{ \rm HO}$) is delayed relative to that driven by pairs ($I_{\rm PW}$), whereas increasing the correlation causes the two processes to synchronize and evolve simultaneously. To further quantify this phenomenon, we define the temporal centroid of $I_{\rm PW}$ and $I_{\rm HO}$ as

\begin{equation}
    \tau_{\rm PW} = \frac{\sum_i t_iI_{\rm PW}(t_i)}{\sum_i I_{\rm PW}(t_i)},  \hspace{1cm}\tau_{\rm HO} = \frac{\sum_i t_iI_{\rm HO}(t_i)}{\sum_i I_{\rm HO}(t_i)}
\end{equation}

The temporal centroid is a measure representing the ``center of mass'' of a distribution in time~\cite{peeters2004large}, commonly used in signal analysis. The difference between the temporal centroids of two signals reflects the relative delay between them, since the temporal centroid corresponds to the average time at which most of the signal occurs. Accordingly, to quantify the delay between infections caused by pairwise and triple interactions, we define
$\Delta \tau = \tau_{\rm PW} - \tau_{\rm HO}$. In particular, $\Delta \tau = 0$ indicates that the two processes evolve synchronously. In Figure~\ref{fig:results3}(c), we plot $\Delta \tau$ as a function of the cross-order hyperdegree correlation $\sigma$ for hypernetworks with increasing variance. In all cases, we observe that the difference between the temporal centroids is larger for anti-correlated hyperdegree distributions. This indicates that $I_{\rm PW}$ and $I_{\rm HO}$ are highly disynchronized for these values, which can be qualitatively observed in Figures~\ref{fig:results3}(i--m). As the cross-order hyperdegree correlation increases, the difference between the temporal centroids decreases, approaching zero. This shows that the two processes evolve more synchronously. We also observe that, although this behavior holds for most cases, hypernetworks with larger variance, and therefore more heterogeneous degree distributions,  tend to exhibit a more pronounced decrease of $\Delta \tau$ in the anti-correlated case.

Based on the previous results, we can formulate some hypotheses regarding the role of cross-order hyperdegree correlations. In particular, from the temporal centroid analysis and the observed changes in the epidemic threshold, we find that when the hyperdegree distributions are strongly correlated, the epidemic onset occurs earlier, and pairwise and group infections develop almost simultaneously. This anticipation effectively lowers the epidemic threshold, as smaller infectivity is sufficient to trigger an outbreak. In contrast, in anti-correlated systems, the infection initially spreads through pairwise interactions and only later activates group-level transmissions.

\subsection{ Hierarchical spread in higher-order networks }\label{sec:results_herarch}

In the previous section, we examined the role of cross-order hyperdegree correlations at two levels: macroscopically, through their effects on the epidemic threshold and bistability region, and microscopically, by analyzing how pairwise and group interactions contribute to infection dynamics over time. Here, we further investigate the role of different node classes in the epidemic process to better understand the mechanisms underlying the previously observed behavior. In particular, we study which nodes are most influential in driving the spreading dynamics across different types of higher-order networks. 

The effective hyperdegree model  provides a natural and practical way to study different hyperdegree classes, i.e., as the equations of the model already aggregate nodes characterized by a specific hyperdegree $(k_1, k_2)$.

We analyze the spreading dynamics through the lens of hierarchical propagation across node classes defined by their hyperdegrees $(k_1, k_2)$. In contagion processes in networks (with only pairwise interactions) with heterogeneous degree distributions, it has been shown that the spreading follows a well-defined hierarchical pattern~\cite{barthelemy2004velocity}. Specifically, the disease initially spreads through highly connected nodes, and once these are infected, it progressively cascades to nodes of smaller  degree.

To extend this analysis to higher-order networks we first look at the evolution of average hyperdegree of newly infected nodes. To quantify this, we define the variable $J_{k_1}^{k_2}(t)$ as the cumulative number of infected nodes in hyperdegree class $(k_1,k_2)$, whose time evolution is described by $\dot{J}_{k_1}^{k_2} = \big(\beta_1 k_1 \braket{I} + \beta_2 k_2 \braket{Z}\big)S_{k_1}^{k_2}$. Given this, we define the average hyperdegree of newly infected nodes for each order of interaction as:

\begin{equation}
    \begin{split}
        \bar{k}_1(t) = \frac{\sum_{k_1} k_1 \bigg(J_{k_1}^{k_2} (t+\Delta t) - J_{k_1}^{k_2}  (t)\bigg)}{J(t+\Delta t ) - J(t)},\\
        \bar{k}_2(t) = \frac{\sum_{k_2} k_2 \bigg(J_{k_1}^{k_2} (t+\Delta t) - J_{k_1}^{k_2}  (t)\bigg)}{J(t+\Delta t ) - J(t)}.
    \end{split}
\end{equation}

\begin{figure}[h!]
    \centering
    \includegraphics[width=0.9\linewidth]{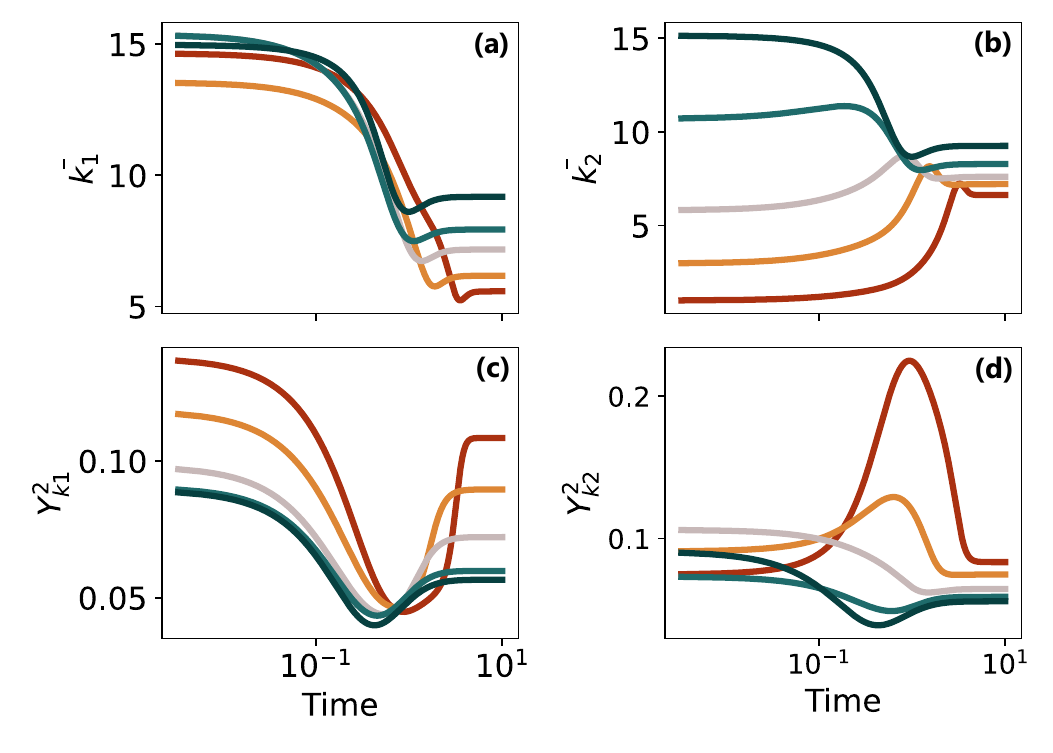}
    \caption{\textbf{Hierarchical Spread in Higher-Order Networks:}Panels (a) and (b) show the temporal evolution of the average 1-hyperedge degree and 2-hyperedge degree, respectively, of newly infected nodes. Panels (c) and (d) display the corresponding time evolution of the participation ratio for pairwise and group interactions, as defined in~\eqref{eq_inv_part}. In all panels, each curve corresponds to a distinct value of the cross-order hyperdegree correlation,  $\sigma = -1, -0.5, 0, 0.5$, and $1$. All hypergraphs have $N = 1000$ nodes with negative binomial hyperdegree distributions for both pairwise and three-body interactions, mean hyperdegrees $\langle k_1 \rangle \approx \langle k_2 \rangle \approx 6$, and variance 30 for both interaction orders. Disease parameters are $\lambda_1 = 0.9$, $\lambda_2 = 3$ and $\gamma=1$. }
\label{fig:results4}
\end{figure}

Here, the values of $J_{k_1}^{k_2} (t)$ are obtained by solving the effective hyperdegree model, and $\Delta t$ is the time step used to solve the system of equations. The quantities $\bar{k}_1(t)$ and $\bar{k}_2(t)$ represent the average hyperdegree of newly infected nodes between time $t$ and $t + \Delta t$, providing insight into the characteristics of newly infected. In Figs.~\ref{fig:results4}(a) and~\ref{fig:results4}(b), we show the temporal evolution of $\bar{k}_1(t)$ and $\bar{k}_2(t)$ for the same cases displayed in panels (i--m) of Fig.~\ref{fig:results3}.

We observe that $\bar{k}_1(t)$ starts at a high value, indicating that nodes infected early in the epidemic tend to be highly connected. As time progresses, $\bar{k}_1(t)$ decreases, reflecting that newly infected nodes have progressively smaller 1-hyperdegree. This behavior is consistent across all correlation types and aligns with what is typically observed in epidemics on networks with just pairwise interaction. It corresponds to a hierarchical spreading pattern in which highly connected nodes are infected at the onset of the epidemic; once these hubs are rapidly infected, the disease spreads to nodes with increasingly smaller values of $k_1$.

However, the evolution of $\bar{k}_2(t)$ varies with the cross-order hyperdegree correlation. In correlated cases, the average 2-hyperdegree exhibits a similar hierarchical pattern to the 1-hyperdegree: highly connected nodes in group interactions are infected first, and as the epidemic progresses, newly infected nodes have smaller 2-hyperdegree. In contrast, in anti-correlated cases we observe the opposite trend: $\bar{k}_2(t)$ starts at a low value and increases as the epidemic progresses, reaching a maximum later in the epidemic. This indicates that anti-correlation alters the traditional hierarchical spreading pattern observed in heterogeneous networks, causing hubs in group interactions to be reached only later in the epidemic.

To further characterize the progression of the disease in this system, we introduce the inverse participation ratio $Y^2_{k_m}(t)$, defined for systems with pairwise interactions in \cite{barthelemy2004velocity}. The inverse participation ratio is given in Eq.~\eqref{eq_inv_part} for both pairwise and  three-body interactions. This quantity measures how strongly the infection is localized within a specific  hyperdegree class at time $t$. A large value of $Y^2_{k_m}(t)$ indicates that the disease is concentrated in a narrow subset of  hyperdegree classes, whereas a small value suggests that the infection is more homogeneously distributed across all  hyperdegree classes.

\begin{equation}
    \begin{split}
        &Y^2_{k_1}(t) = \sum_{k_1} \bigg(\frac{\sum_{k_2}I^{k_2}_{k_1}(t)}{I(t)}\bigg)^2 \\
        &Y^2_{k_2}(t) = \sum_{k_2} \bigg(\frac{\sum_{k_1}I^{k_2}_{k_1}(t)}{I(t)}\bigg)^2
    \end{split}
    \label{eq_inv_part}
\end{equation}

In Fig.~\ref{fig:results4}(c) and \ref{fig:results4}(d) we show the temporal evolution of the inverse participation ratio for the cases displayed in panels (i–m) of Fig.~\ref{fig:results3}. Similar to the behavior of the average hyperdegree of newly infected nodes, we observe that the dynamics under pairwise interactions are similar across different cross-hyperdegree correlation types. Specifically, \(Y^2_{k_1}(t)\) is initially large, indicating that the disease is highly localized within specific hyperdegree classes, likely the highly connected nodes, as also suggested by the evolution of \(\bar{k}_1(t)\). As time progresses, the infection becomes more homogeneously distributed, before becoming slightly localized again once the epidemic reaches the larger population of low-degree nodes.

The evolution of \(Y^2_{k_2}(t)\), however, displays a clear dependence on the cross-order hyperdegree correlation. For positive correlations between interaction orders, the disease is again localized in specific hyperdegree classes at the beginning of the outbreak, followed by a phase where it becomes more evenly distributed. In contrast, when the hyperdegrees of different interaction orders are anti-correlated, a strong localization emerges at later times, approximately when the inverse participation ratio for pairwise interactions reaches its minimum. This indicates that when the disease is homogeneously spread across the 1-hyperdegree classes, it becomes highly localized within the 2-hyperdegree classes. Comparing this with the evolution of \(\bar{k}_2(t)\), we see that this strong localization occurs when the average hyperdegree in three-body interactions reaches its peak.

These findings highlight how the role of different hyperdegree classes changes under different cross-order hyperdegree correlations. It is worth noting that this description applies when the number of initially infected nodes is small; when the initial proportion is large, higher-order infections do not require pairwise-driven early spreading.

\subsection{Control of influential nodes}\label{sec:results_control}

Building on the results presented in the previous section, a natural next step is to investigate the impact of targeted control strategies applied to specific hyperdegree classes. In this context, we define a control strategy as a mitigation intervention in which a selected subset of nodes is made immune to the disease, with the goal of reducing the overall epidemic spread in the system \cite{hadjichrysanthou2015epidemic}. Classical control methods often choose this subset at random \cite{anderson1991infectious}, but for spreading processes on heterogeneous networks it has been shown that targeted selection based on centrality measures typically outperforms random immunization \cite{hadjichrysanthou2015epidemic, callaway2000network}.

The role of influential nodes in hypergraphs has been explored in previous work. For example, the authors in \cite{st2022influential} showed that nodes with high hyperdegree are particularly effective as epidemic seeds, and that influential groups can dominate the spreading dynamics when non-linear contagion is present. As demonstrated in the previous section, cross-order hyperdegree correlations can significantly alter the role that different hyperdegree classes play throughout the infection process.

To develop our control scheme, we assign to each node $i$ a connectivity weight $ w_i(\alpha) = (1 - \alpha) k_{1}^i + \alpha k_{2}^i$, where $k_{1,i}$ and $k_{2,i}$ are the 1- and 2-hyperdegrees of node $i$, and $\alpha$ is a control parameter that tunes the relative importance of pairwise versus group connectivity. We then rank all nodes according to $w_i(\alpha)$ and select the top 5\% with the highest weight. These selected nodes are assigned a recovery rate 100 times larger than that of the rest of the population, i.e., $\gamma_v = 100 \gamma$. In practice, the targeted nodes recover almost instantaneously after infection and therefore effectively do not participate in transmitting the disease to other nodes. Note that when $\alpha = 0$, the weight reduces to $w_i = k_1^i$, so the controlled nodes correspond to those with the largest 1-hyperdegree; similarly, when $\alpha = 1$, the top 5\% correspond to nodes that participate in the largest number of three-body interactions.

\begin{figure*}[ht!]
    \centering
    \includegraphics[width=0.9\linewidth]{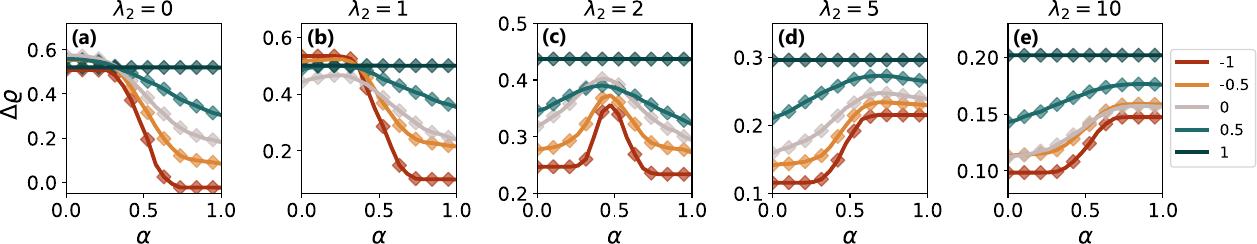}
    \caption{\textbf{Control of influential nodes:} Panels (a–e) depict the prevalence reduction, $\Delta\varrho$, as a function of the control parameter $\alpha$ for $\lambda_2 = 0, 1, 2, 5,$ and $10$. The results are obtained from solutions of the compact effective hyperdegree model. In all panels, each curve corresponds to a distinct value of the cross-order hyperdegree correlation, $\sigma = -1, -0.5, 0, 0.5$ and  $1$. All hypergraphs have negative binomial hyperdegree distributions for both pairwise and three-body interactions with mean hyperdegrees $\langle k_1 \rangle \approx \langle k_2 \rangle \approx 6$ and variance 30 for both interaction orders. The infection rate for pairwise interactions is $\lambda_1 = 0.9$. }
\label{fig:results5}
\end{figure*}

To evaluate the effectiveness of this control mechanism, we applied it to the hypernetworks shown in Figs.~\ref{fig:results3}(i--m) and tested the scheme for different values of $\alpha$. As a baseline for comparison, we also simulated the uncontrolled case in which no nodes are targeted. For each parameter combination, we computed the reduction in the final epidemic size $\Delta \varrho = (\rho_b - \rho_c)/\rho_b$, where $\rho_b$ is the epidemic size in the uncontrolled system and $\rho_c$ is the epidemic size obtained when controlling the top 5\% of nodes with the highest $w_i(\alpha)$. We repeat this procedure for different values of $\lambda_2$. Results are shown in figure \ref{fig:results5}.

For correlated hypernetworks, corresponding to the case $\sigma = 1$, we observe that the reduction in final epidemic size is constant and independent of the value of $\alpha$. This is expected, since in this fully correlated scenario the highly connected nodes in pairwise interactions are the same nodes that are highly connected in group interactions. Furthermore, as described in the previous section, the roles of the different hyperdegree classes are qualitatively the same for both orders of interaction in this scenario.

In contrast, for $\sigma=-1$ we see an inversion in the role of $\alpha$. For low values of $\lambda_2$ (shown in Figs.~\ref{fig:results5}a and \ref{fig:results5}b), the largest reduction in epidemic size occurs for $\alpha \to 0$, meaning that targeting nodes with high pairwise connectivity is most effective. For larger values of $\lambda_2$, as shown in Figs.~\ref{fig:results5}d and \ref{fig:results5}e, the largest reduction occurs for $\alpha \to 1$, where the controlled nodes are those with high connectivity in group interactions. Finally, for intermediate values of $\lambda_2$, as observed in Fig ~\ref{fig:results5}c,  we see that the maximum reduction appears around $\alpha \approx 0.5$, indicating that the optimal control strategy is not necessarily to target the highest-degree nodes in either 1- or 2-hyperdegree alone, but rather those with the largest overall number of neighbors across both types of interactions. This effect arises when pairwise and three-body interactions contribute comparably to the epidemic.

Furthermore, the prevalence reduction in the uncorrelated case exhibits a similar trend to the anti-correlated case, but it is of a different magnitude. This is because, in both configurations, there exists a subset of nodes that are highly connected in one order of interaction but not in another. As a result, varying $\alpha$ produces a comparable effect on $\Delta \varrho$, though the magnitude is larger in the anti-correlated case, where all highly connected nodes participate exclusively in a single order of interaction. This behavior also extends to the moderately correlated regime ($\sigma = 0.5$), where a smaller subset of such order-exclusive highly connected nodes exists. These results confirm that the inversion in the roles of nodes, regarding both which nodes drive the spreading process and are most efficient to control, is a general consequence of how cross-order hyperdegree correlations alter the role of highly connected nodes in epidemic dynamics.


\label{sec:conclusion}

\section{Discussion and Conclusions}

The study of contagion processes on higher-order networks has typically assumed that hyperdegrees at different interaction orders are independently distributed. While analytically convenient, this assumption obscures a potentially crucial structural feature: in real systems, a node's participation in pairwise interactions may correlate with its participation in group interactions. In this paper, we developed a tractable framework that explicitly captures these cross-order hyperdegree correlations and reveals their profound impact on epidemic dynamics.

Our effective hyperdegree model accurately reproduces stochastic simulations across diverse correlation regimes (Fig.~\ref{fig:results1}), validating its use for systematic analysis. The model reveals that cross-order hyperdegree correlations influence spreading dynamics through modification of hub identity and temporal coordination of transmission pathways. Positive cross-order hyperdegree correlations anticipate the epidemic onset, lowering the epidemic threshold (Fig.~\ref{fig:results3}a,b), as nodes that are highly connected through pairwise interactions simultaneously serve as hubs for group interactions. When infection reaches these highly connected nodes, it gains simultaneous access to both transmission pathways, creating parallel routes for propagation. In anti-correlated systems, hubs are specialized, pairwise hubs have few group connections and vice versa, requiring sequential activation of different hub classes and thus raising the threshold.

This hub specialization produces temporal desynchronization of spreading pathways. Our analysis of temporal centroids (Fig.~\ref{fig:results3}c) shows that in anti-correlated systems, pairwise infections ($I_{PW}$) occur significantly earlier than group infections ($I_{HO}$). The infection spreads first through pairwise hubs, then percolates through medium-degree nodes before activating group hubs. In contrast, positively correlated systems exhibit synchronized growth of both transmission modes from outbreak onset. 

Classical results establish that epidemics on heterogeneous networks follow a hierarchical pattern: high-degree nodes infect first, then the disease cascades to lower-degree classes. We find this pattern holds for pairwise interactions regardless of correlation structure---$\bar{k}_1(t)$ consistently decreases over time. However, cross-order hyperdegree correlations produce a qualitative inversion for group interactions. In positively correlated systems, $\bar{k}_2(t)$ exhibits the standard decreasing pattern. In anti-correlated systems, $\bar{k}_2(t)$ increases during early epidemic phases (Fig.~\ref{fig:results4}), reflecting delayed activation of group hubs. The inverse participation ratio $Y^2_{k_2}$ confirms this reorganization (Fig.~\ref{fig:results4}d): in anti-correlated systems, $Y^2_{k_2}$ peaks at intermediate times when pairwise spreading has homogenized across degree classes, indicating strong localization within specific 2-hyperdegree classes. This two-phase dynamics,  initial pairwise-dominated spreading followed by group amplification, illustrates how cross-order correlations can reshape the spreading hierarchy in higher-order systems.

Cross-order hyperdegree correlations also have direct consequences for epidemic control. Our targeted intervention scheme (Fig.~6) reveals that the optimal balance between pairwise and group targeting is dictated by the correlation structure: in fully correlated systems, the choice is irrelevant since the same nodes dominate both interaction orders, whereas in anti-correlated and uncorrelated systems, the optimal strategy shifts depending on the relative strength of group transmission. These results show that effective interventions require knowledge of correlation structure, not just hyperdegree distributions. More broadly, this highlights that control strategies in higher-order networks are not a straightforward extension of pairwise spreading with added infection pathways. Instead, the interplay between interaction orders dictates which nodes are most effective to target, making pairwise-based intuition insufficient.

In summary, we demonstrate that cross-order hyperdegree correlations fundamentally reshape epidemic spreading through three mechanisms: (i) positive correlations lower epidemic thresholds by creating redundant transmission pathways through nodes serving as hubs for multiple interaction orders; (ii) anti-correlations temporally desynchronize pairwise and group transmission, with pairwise spreading preceding group amplification; (iii) anti-correlations invert the traditional hierarchical spreading patterns for group interactions. These effects have direct implications for epidemic control: optimal targeting strategies depend critically on both correlation structure and relative transmission strengths, with anti-correlated systems requiring adaptive strategies that shift between pairwise and group targeting as epidemics evolve.

This analysis could be extended to hypernetworks with more than two orders of interaction, where the cross-order correlations are described by a correlation matrix. We expect some of the behaviour described in this paper to carry over to such systems. For instance, positively correlated hyperdegree distributions would synchronise infections across all orders, while uncorrelated distributions would lead to pairwise-driven early spreading followed by subsequent bursts as hubs in other orders are reached. Moreover, such systems could exhibit a mixture of positively and negatively correlated orders, where nodes are hubs in some orders but peripheral in others, giving rise to more complex hierarchical structures and competing roles within the network. Overall, the full correlation matrix would shape the effectiveness of control strategies, as the choice of which nodes to target would depend on the complete pattern of correlations across orders. While the higher-order network generating model can be easily extended to accommodate additional orders of interaction, the effective hyperdegree epidemic model would gain considerable complexity, as the number of hyperedge arrangement states grows rapidly with each additional order.

More broadly, our results suggest that the interplay of higher-order interactions, not merely their presence, plays an important role in driving dynamics in hypernetworks. As empirical studies increasingly reveal higher-order interactions in social, biological, and technological systems, accounting for cross-order correlations may prove valuable for predictive modeling and the design of effective interventions. Natural directions for future work include developing methods to measure cross-order hyperdegree correlations from empirical data, extending the framework to other dynamical processes, and exploring how correlation structure interacts with other structural features.

\section*{Acknowledgments}
A.G. acknowledges the PhD studentship support from Northeastern University London. F.M. acknowledges support from the Austrian Science Fund (FWF) through project 10.55776/PAT1652425.

\section*{Data availability} The data that support the findings of
this Letter are openly available \cite{andresgithub}

\clearpage

\bibliographystyle{unsrt}
\bibliography{ref}

\clearpage
\onecolumngrid
\section*{Supplementary material of \emph{Unveiling the impact of cross-order hyperdegree correlations in contagion processes on hypergraphs}}
This Supplementary Material provides additional details on the methods and models used in the paper \textit{Unveiling the Impact of Cross-Order Hyperdegree Correlations in Contagion Processes on Hypergraphs}. It is organized as follows. First, in section \ref{conf_model_SM} we present a detailed description of the configuration model employed to generate higher-order networks with varying levels of heterogeneity and tunable cross-order hyperdegree correlations. Section~\ref{negbin} describes how the negative binomial distribution was parametrized to generate hypergraphs with heterogeneous hyperdegree distributions. Then, in Sections \ref{sec:SIS_effdegree_SM} and \ref{sec:SIS_compeffdegree_SM}, we provide the full mathematical derivation of the effective hyperdegree model and its compact version, respectively. Finally, in \ref{sec:dimentionality_SM}, we show the dimensionality reduction in the number of equations obtained from the compact effective hyperdegree model.

\subsection*{Configurational model} \label{conf_model_SM}

We define a hypergraph $H = (\mathcal{N}, \mathcal{E})$, where $\mathcal{N}$ is the set of $N = |\mathcal{N}|$ nodes and $\mathcal{E}$ is the set of $E = |\mathcal{E}|$ hyperedges representing their interactions. Each hyperedge $e \in \mathcal{E}$ is a subset of $\mathcal{N}$ and can be characterized by its order $m = |e| - 1$. For example, $m = 1$ corresponds to pairwise interactions, $m = 2$ corresponds to group interactions of three nodes, and so on. We denote the maximum order of interaction in the system as $M$.

A generic node $i$ has $M$ hyperdegrees, denoted by $k_m^{i}$, which represent the number of interactions of order $m$ (i.e., $m$-hyperedges) incident to node $i$. The hyperdegrees are distributed according to a joint probability distribution $\mathbf{P}(\mathbf{k})$, which gives the probability that a randomly chosen node has hyperdegree vector $\mathbf{k} = [k_1, k_2, \ldots, k_M].$

We aim to develop a random hypergraph model designed to generate higher-order networks in which hyperdegree vectors are drawn from a distribution $\mathbf{P}(\mathbf{k})$, allowing arbitrary marginal distributions and controlled correlations between different interaction orders. We define these correlations as \emph{cross-order degree correlations}, which quantify how a node's participation in hyperedges of one order relates to its participation in hyperedges of another order.

For a network with $M$ interaction orders, the cross-order degree correlations are captured by the correlation matrix
\begin{equation}
\mathbf{\Sigma} =
\begin{bmatrix}
1 & \sigma_{1,2} & \sigma_{1,3} & \cdots & \sigma_{1,M} \\
\sigma_{2,1} & 1 & \sigma_{2,3} & \cdots & \sigma_{2,M} \\
\vdots & \vdots & \vdots & \ddots & \vdots \\
\sigma_{M,1} & \sigma_{M,2} & \sigma_{M,3} & \cdots & 1
\end{bmatrix}.
\end{equation}

Here, the element $\sigma_{m,m'}$ quantifies the Pearson correlation coefficient between the list of $m$-hyperdegrees $(k_m^1, k_m^2, \ldots, k_m^N)$ and the list of $m'$-hyperdegrees $(k_{m'}^1, k_{m'}^2, \ldots, k_{m'}^N)$ across all nodes. Note that $\sigma_{m,m'} = \sigma_{m',m}$.

As input to our hypergraph generator, we require the desired marginal distributions $P_1(k_1), P_2(k_2), \ldots, P_M(k_M)$ and the matrix of cross-order hyperdegree correlations $\mathbf{\Sigma}$. The algorithm used to generate the hyperdegree vector for each node is detailed below:

\begin{enumerate}
    \item First, we generate samples with the desired cross-order degree correlations by drawing $N$ vectors from a multivariate normal distribution $\mathcal{N}_M(\boldsymbol{\mu}, \boldsymbol{\Sigma})$, where $\boldsymbol{\mu}$ is the mean vector (typically set to $\mathbf{0}$ as it will be transformed away in subsequent steps) and $\boldsymbol{\Sigma}$ is the covariance matrix encoding the desired correlations. For each node $n = 1, 2, \ldots, N$, we obtain a sample vector $(z_1^n, z_2^n, \ldots, z_M^n)$.
    
    \item From the samples obtained in the previous step, we extract the hyperdegree lists for each interaction order. Specifically, for each order $m = 1, 2, \ldots, M$, we collect the $m$-th component from all hyperdegree vector sample to form the list $(z_m^1, z_m^2, \ldots, z_m^N)$. Each of these lists corresponds to samples from a univariate normal distribution with marginal variance determined by $\boldsymbol{\Sigma}$.
    
    \item For each node's hyperdegree vector, we apply the univariate standard normal CDF $\Phi(\cdot)$ to each component separately. That is, for the $n$-th node with sampled values $(z_1^n, z_2^n, \ldots, z_M^n)$ from step 1, we compute $(u_1^n, u_2^n, \ldots, u_M^n) = (\Phi(z_1^n), \Phi(z_2^n), \ldots, \Phi(z_M^n))$, where each $u_m^n \in (0,1)$. This transformation yields uniformly distributed marginals while preserving the correlation structure.
    
    \item For each interaction order $m = 1, 2, \ldots, M$, we apply inverse transform sampling to convert the uniform samples to the desired marginal distribution. Specifically, for each node $n$, we compute the hyperdegree $k_m^n = F_m^{-1}(u_m^n)$, where $F_m^{-1}$ is the inverse cumulative distribution function (quantile function) of the desired marginal distribution $P_m(k_m)$. This procedure is repeated for all interaction orders and all nodes, yielding a complete hyperdegree vector $\mathbf{k}^n = (k_1^n, k_2^n, \ldots, k_M^n)$ for each node $n$.
    
    \item The resulting hyperdegree vectors preserve the original correlations between interaction orders encoded in $\boldsymbol{\Sigma}$, while the marginal distribution of each order follows the desired $P_m(k_m)$. This is guaranteed by the properties of the Gaussian copula construction~\cite{devroye2006nonuniform}.
\end{enumerate}

A limitation of this method is that the functional form of each desired marginal distribution $P_m(k_m)$ must be known, along with its cumulative distribution function $F_m(k_m)$, which must be invertible. In practice, many commonly used distributions (e.g., Poisson, power-law, exponential) satisfy these requirements, and numerical methods can be employed for distributions without closed-form CDFs.

Using this algorithm, we obtain a hyperdegree vector for each node in the system. We assign one hyperdegree vector to each node and generate hyperstubs---half-edges or unconnected elements---for each interaction order $m$ according to the sampled hyperdegrees. For each order $m$, we randomly group $m+1$ hyperstubs to form hyperedges of that order, repeating this process until all hyperstubs of order $m$ are exhausted.

With this formulation, we can generate hypergraphs with arbitrary marginal degree distributions while incorporating tunable cross-order degree correlations.

It is worth noting that in this formulation we do not explicitly constrain the existence of nested links. Nestedness has been shown to profoundly shape spreading processes in complex networks \cite{malizia2025disentangling, lamata2025hyperedge, kim2024higher}. However, for the degree distributions used in this paper, the probability of having nested links within three-node interactions is very small. To ensure that this structural property did not confound our exploration of the desired features, we calculated the inter-order overlap, as defined in \cite{lamata2025hyperedge}, for all higher-order networks used in this study. All values of inter-order overlap were below 0.05, indicating that of all possible nested links, only 5\% or fewer actually existed in the generated networks. The dynamical effect of this quantity is therefore negligible, allowing us to isolate the influence of heterogeneity and cross-order correlations without significant confounding from nestedness effects. Furthermore, in a fully nested system (such as a simplicial complex formulation), there is no way to disentangle the hyperdegree distributions, and as a consequence, the hyperdegrees are always positively correlated. Our configuration model approach avoids this constraint, enabling independent control of marginal distributions and cross-order correlations.

\subsection*{Hypergraphs with heterogeneous hyperdegree distributions}\label{negbin}

Using the configuration model framework described in Section~\ref{conf_model_SM}, we construct higher-order networks with pairwise interactions ($m=1$) and three-body interactions ($m=2$). Each node $i$ is characterized by its hyperdegree vector $(k_1^{i}, k_2^{i})$, where $k_1^{i}$ denotes the number of pairwise connections and $k_2^{i}$ denotes the number of three-body interactions in which node $i$ participates.

We specify marginal hyperdegree distributions using negative binomial (NB) distributions:
\begin{equation}
    P(k_1) \sim \text{NB}(r_1, p_1) \quad \text{and} \quad P(k_2) \sim \text{NB}(r_2, p_2)
    \label{eq:nb_traditional}
\end{equation}
The negative binomial distribution is well-suited for modeling complex networks as it captures heterogeneity ranging from Poisson-like (low variance) to highly skewed distributions (high variance). We reparametrize in terms of mean and variance: $\text{NB}(\langle k_m \rangle, \text{var}_m)$, with
\begin{equation}
    r_m = \frac{\langle k_m \rangle^2}{\text{var}_m - \langle k_m \rangle}, \qquad 
    p_m = \frac{\langle k_m \rangle}{\text{var}_m}
    \label{eq:nb_reparametrization}
\end{equation}
valid when $\text{var}_m > \langle k_m \rangle$. This parametrization allows systematic control of network heterogeneity by varying $\text{var}_m$ while fixing $\langle k_m \rangle$. When $\text{var}_m \approx \langle k_m \rangle$, the distribution is relatively homogeneous (similar to a Poisson distribution), with most nodes having hyperdegrees close to the mean. As $\text{var}_m$ increases while keeping $\langle k_m \rangle$ constant, the distribution becomes increasingly skewed, with a small number of high-degree hub nodes and many low-degree nodes. In our simulations, we typically maintain small average hyperdegrees ($\langle k_1 \rangle, \langle k_2 \rangle \in [2, 10]$), while varying $\text{var}_m$ to explore different heterogeneity regimes.

Finally, we note that our formulation does not explicitly constrain the existence of nested links, which have been shown to profoundly shape spreading processes in complex networks~\cite{malizia2025disentangling, lamata2025hyperedge, kim2024higher}. However, for the degree distributions used in this paper, the probability of having nested links within three-body interactions is very small.

\subsection*{SIS Higher-order effective degree model, full derivation} 
\label{sec:SIS_effdegree_SM}

We consider a Susceptible-Infected-Susceptible (SIS) epidemic process on hypergraphs, where the transmission mechanism explicitly accounts for higher-order interactions. In this framework, infection of a susceptible node occurs when it belongs to an $m$-hyperedge (a hyperedge connecting $m$ nodes) in which all other $m-1$ nodes are simultaneously infected. Each interaction order $m$ is characterized by its own infection rate $\beta_m$, reflecting the potentially different transmission dynamics at different orders of interaction. The recovery rate $\gamma$, in contrast, is assumed to be independent of network structure and applies uniformly across all infected nodes. Throughout this work, we focus on a system with two distinct interaction orders: $m=1$, representing standard pairwise (dyadic) interactions between two nodes, and $m=2$, representing three-body (triadic) interactions. While our analysis centers on these two orders, the mathematical framework we develop is readily extensible to accommodate higher-order interactions with $m \geq 3$.

The model introduced here represents a generalization of the effective-degree framework originally developed for network-based epidemics with pairwise interactions only \cite{lindquist2011effective}. In the classical effective degree approach, the system state is tracked through variables such as $S_{s,i}$, which count the number of susceptible nodes having $s$ links to other susceptible nodes and $i$ links to infected nodes. This reduction in dimensionality, from tracking each individual node to tracking groups of nodes with equivalent local environments, enables tractable yet accurate analysis of epidemic dynamics on complex networks. Our extension to higher-order interactions requires a fundamental expansion of this state-space representation. Rather than characterizing nodes solely by their pairwise connections, we must account for all possible configurations of the different types of hyperedges incident to each node. This approach shares conceptual similarities with the composite degree framework introduced by Chen et al. \cite{chen2023composite}, though their work was developed in a discrete-time setting, whereas our formulation operates in continuous time.

The key innovation in our framework is the characterization of each node by a comprehensive \textbf{neighborhood vector} that captures the complete local infection landscape. Consider a randomly selected node $u$ in the hypergraph. The state of all nodes connected to $u$ through hyperedges of different orders is described by the five-dimensional vector:
\begin{equation}
\mathbf{n}_u = [s, i, x, y, z]
\end{equation}
where each component has a precise epidemiological interpretation:
\begin{itemize}
    \item \textbf{$s$}: number of pairwise links (1-hyperedges) connecting $u$ to susceptible nodes
    \item \textbf{$i$}: number of pairwise links connecting $u$ to infected nodes  
    \item \textbf{$x$}: number of three-body interactions (2-hyperedges) in which $u$ participates alongside two other susceptible nodes
    \item \textbf{$y$}: number of three-body interactions in which $u$ participates with one susceptible and one infected node
    \item \textbf{$z$}: number of three-body interactions in which $u$ participates with two infected nodes
\end{itemize}
This vector provides a complete local description of the infection pressure experienced by node $u$ and determines both its instantaneous infection risk (if susceptible) and its potential to transmit infection (if infected).

Rather than tracking each individual node, we exploit the symmetry of nodes sharing identical neighborhood vectors. We define:
\begin{itemize}
    \item \textbf{$S_{s,i}^{x,y,z}(t)$}: the total number of susceptible nodes with neighborhood vector $\mathbf{n}_u = [s,i,x,y,z]$ at time $t$
    \item \textbf{$I_{s,i}^{x,y,z}(t)$}: the total number of infected nodes with the same neighborhood vector
\end{itemize}
These aggregate variables form the foundation of our dynamical system. The state space dimension grows with the maximum degree in each interaction order, but remains vastly smaller than the full individual-based representation for large networks.

Beyond node-centric variables, we introduce hyperedge-centric quantities that track the infection states within individual hyperedges. These variables provide an alternative perspective on the system state that proves essential for closing the dynamical equations.

For pairwise interactions ($m=1$), we denote by $[AB]$ the total number of 1-hyperedges (links) connecting one node in state $A$ to one node in state $B$, where $A, B \in \{S, I\}$. For example:
\begin{itemize}
    \item $[SS]$: number of links between two susceptible nodes
    \item $[SI]$ (equivalently $[IS]$): number of links between a susceptible and an infected node  
    \item $[II]$: number of links between two infected nodes
\end{itemize}
For three-body interactions ($m=2$), we use the notation $[ABC]$ to denote the total number of 2-hyperedges (three-node hyperedges) containing nodes in states $A$, $B$, and $C$. Examples include:
\begin{itemize}
    \item $[SSS]$: three-body interactions with three susceptible nodes
    \item $[SSI]$ (equivalently $[SIS]$ or $[ISS]$): three-body interactions with two susceptible and one infected node
    \item $[SII]$ (equivalently $[ISI]$ or $[IIS]$): three-body interactions with one susceptible and two infected nodes  
    \item $[III]$: three-body interactions with three infected nodes
\end{itemize}

In our model we also track \textbf{hyperedge arrangements}: configurations formed by two hyperedges of different orders sharing a common node. These structures capture the correlation between a node's status in different types of hyperedges and prove essential for accurately modeling neighborhood dynamics.

We denote such structures using notation of the form $A\underline{B}CD$, where:
\begin{itemize}
    \item The \textbf{underlined symbol} $\underline{B}$ indicates the shared node
    \item The symbol to the \textbf{left} of the underline ($A$) represents the state of the node connected to $\underline{B}$ via a 1-hyperedge
    \item The symbols to the \textbf{right} of the underline ($C$ and $D$) represent the states of the two nodes connected to $\underline{B}$ via a 2-hyperedge
\end{itemize}
For example, $I\underline{S}II$ represents a configuration where: (1) a susceptible node (the underlined $S$) participates in a pairwise link with an infected node ($I$), and (2) the same susceptible node simultaneously participates in a three-body interaction with two infected nodes ($II$).

\begin{figure}[h!]
    \centering
    \includegraphics[width=0.9\linewidth]{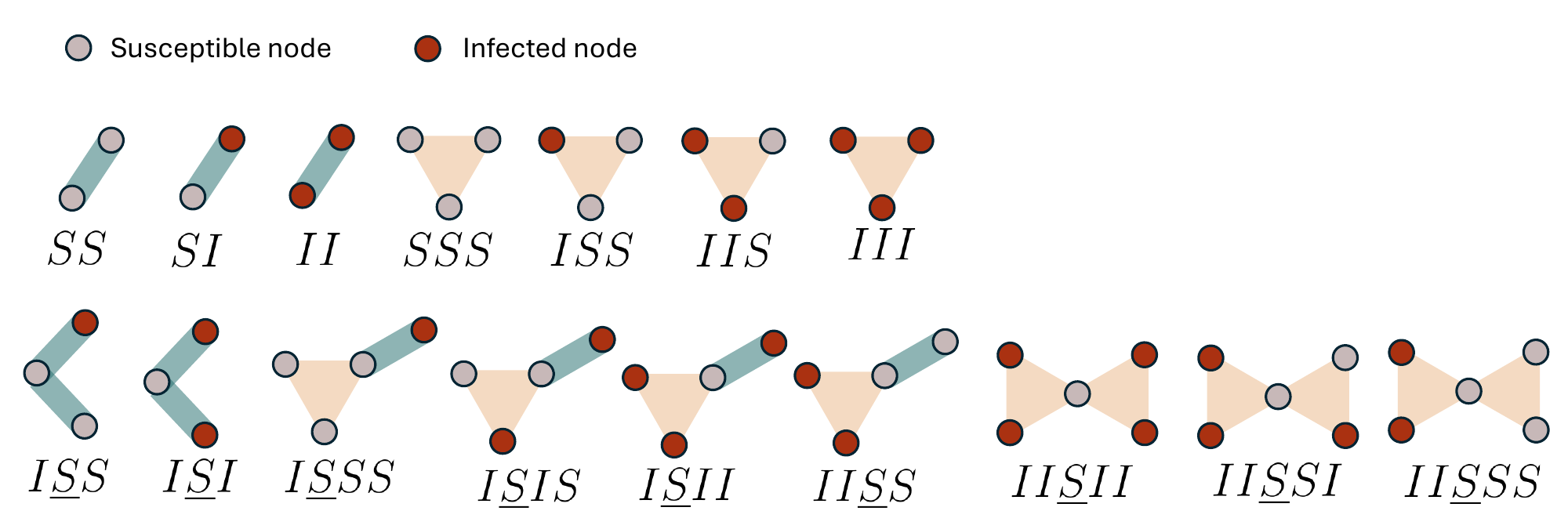}
    \caption{\textbf{Hyperedge and hyperedge arrangement states}: In the first row, we show representations of different configurations that hyperedges can take, considering the states of the nodes within each hyperedge. The second row shows hyperedge arrangements composed of two hyperedges sharing a common node. Only the hyperedge arrangements required for the equations of the effective hyperdegree model are displayed. Grey nodes represent healthy individuals, while red nodes represent infected individuals.} 
    \label{fig:motifs}
\end{figure}

These composite motifs are illustrated comprehensively in Figure~\ref{fig:motifs}, which provides visual representations of all relevant configurations.

To connect the evolution of neighbourhood vectors with the dynamics of hyperedge configurations, we introduce a set of joint variables that simultaneously track both the state of a focal node and the specific local motif structure it participates in. We define $[A\underline{B}CD_{s,i}^{x,y,z}]$ as the number of nodes in state $D$ with neighborhood vector $\mathbf{n}_u = [s,i,x,y,z]$ that participate in a composite motif of type $A\underline{B}CD$. More specifically, this counts nodes in state $D$ that:
\begin{enumerate}
    \item Belong to a 2-hyperedge with two other nodes in states $C$ and $B$
    \item Have the node in state $B$ also connected via a 1-hyperedge to a node in state $A$
    \item Have overall neighborhood configuration $[s,i,x,y,z]$
\end{enumerate}
\textbf{Example}: $[I\underline{S}SS_{s,i}^{x,y,z}]$ counts susceptible nodes with neighborhood state $\mathbf{n}_u = [s,i,x,y,z]$ that belong to a three-body interaction with two other susceptible nodes, where one of these susceptible neighbors is also connected via a pairwise link to an infected node.

These joint variables encode the fine-grained information necessary to determine how transitions in one part of the network propagate to affect neighborhood vectors elsewhere in the system. The analytical tractability of the effective degree framework rests on a critical \textbf{mean-field approximation}: all neighbors of nodes in a given state are assumed to be statistically equivalent. Specifically, we assume that a neighbor of a susceptible node has the same probability of participating in additional hyperedges (of any type) as any other neighbor of any other susceptible node in the network. Under this assumption, the microscopic details of network topology are summarized by aggregate quantities, and the joint variables can be approximated using ratios of global hyperedge counts.

\begin{figure}[h!]
    \centering
    \includegraphics[width=0.8\linewidth]{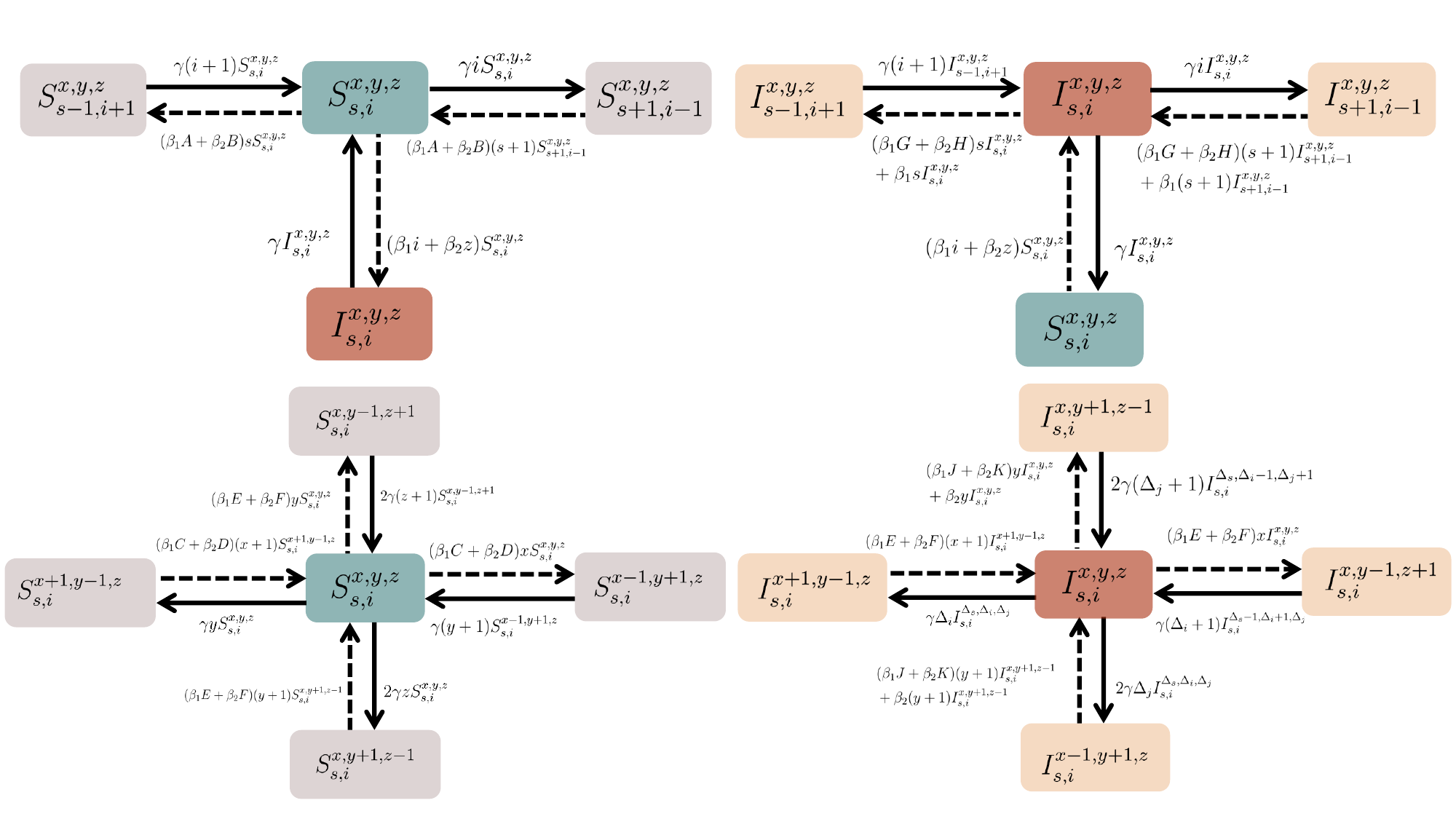}
    \caption{\textbf{State transition diagram:} Diagram showing all possible transitions of the variables $S_{s,i}^{x,y,z}$ and $I_{s,i}^{x,y,z}$ to other states. In the top row, we show transitions due to infection or recovery of the focal node, as well as transitions caused by infection or recovery of neighbors connected to the focal node through pairwise interactions. In the second row, we show transitions resulting from infection or recovery of neighbors connected to the focal node through three-body interactions. Dashed arrows indicate transitions due to infection, while solid arrows indicate transitions due to recovery.}
    \label{fig:transitions_2}
\end{figure}

For example, the joint variable counting specific configurations can be approximated as:
\begin{equation}
[I\underline{S}SS_{s,i}^{x,y,z}] \approx \frac{[I\underline{S}SS]}{[SSS]} \cdot x S_{s,i}^{x,y,z}
\end{equation}
This same logic extends to infected nodes and to all other combinations of states and motif structures, enabling us to express all joint variables in terms of the fundamental state variables $S_{s,i}^{x,y,z}$, $I_{s,i}^{x,y,z}$, and global motif counts. The evolution of $S_{s,i}^{x,y,z}(t)$ and $I_{s,i}^{x,y,z}(t)$ is governed by multiple simultaneous processes that alter either the infection state of the focal node or the composition of its neighborhood. Consider a susceptible node $u$ with neighborhood vector $\mathbf{n}_u = [s,i,x,y,z]$. This node's classification can change through the following mechanisms:
\begin{enumerate}
    \item \textbf{Direct infection of the focal node}: Node $u$ becomes infected through contact with infected neighbors via either pairwise links (at rate $\beta_1 i$) or three-body interactions (at rate $\beta_2 z$), moving from $S_{s,i}^{x,y,z}$ to $I_{s,i}^{x,y,z}$

    \item \textbf{Neighborhood transitions due to neighbor infections}: When one of $u$'s susceptible neighbors becomes infected:
    \begin{itemize}
        \item A susceptible neighbor connected via a pairwise link transitions the node from $S_{s,i}^{x,y,z}$ to $S_{s-1,i+1}^{x,y,z}$
        \item A susceptible neighbor in a three-body interaction (of type $[SSS]$) transitions the node from $S_{s,i}^{x,y,z}$ to $S_{s,i}^{x-1,y+1,z}$
        \item A susceptible neighbor in a mixed three-body interaction (of type $[ISS]$) transitions the node from $S_{s,i}^{x,y,z}$ to $S_{s,i}^{x,y-1,z+1}$
    \end{itemize}

    \item \textbf{Neighborhood transitions due to neighbor recoveries}: When one of $u$'s infected neighbors recovers (at rate $\gamma$), the reverse transitions occur
\end{enumerate}
The complete set of transition pathways is illustrated in Figure~\ref{fig:transitions_2}, which provides a visual representation of all possible state changes and their rates.

Combining all transition mechanisms and applying the mean-field approximations, we obtain the following system of ordinary differential equations governing the evolution of our state variables:
\begin{equation}
    \begin{split}
        \dot{S}_{s,i}^{x,y, z} = & -(\beta_1i+\beta_2z) S_{s,i}^{x,y,z} + \gamma I_{s,i}^{x,y,z}
        \\ & + \Big(\beta_1A+\beta_2B\Big)\Big( (s+1)S_{s+1,i-1}^{x,y,z} -s S_{s,i}^{x,y,z} \Big)
        \\ & + \Big(\beta_1C+\beta_2D\Big)\Big( (x+1)S_{s,i}^{x+1,y-1,z} -x S_{s,i}^{x,y,z} \Big)
        \\ & + \Big(\beta_1E+\beta_2F\Big)\Big( (y+1)S_{s,i}^{x,y+1,z-1} -y S_{s,i}^{x,y,z} \Big)
        \\ & + \gamma \Big(-(i+y+2z)S_{s,i}^{x,y,z} + (i+1)S_{s-1,i+1}^{x,y,z} \\
        & \qquad\qquad + (y+1)S_{s,i}^{x-1,y+1,z} + 2(z+1)S_{s,i}^{x,y-1,z+1} \Big) \\[1em]
        \dot{I}_{s,i}^{x,y, z} = & (\beta_1i+\beta_2z) S_{s,i}^{x,y,z} - \gamma I_{s,i}^{x,y,z}
        \\ & + \Big(\beta_1+ \beta_1G+\beta_2H\Big)\Big( (s+1)I_{s+1,i-1}^{x,y,z} -s I_{s,i}^{x,y,z} \Big) 
        \\ & + \Big(\beta_1E+\beta_2F\Big)\Big( (x+1)I_{s,i}^{x+1,y-1,z} -x I_{s,i}^{x,y,z} \Big)
        \\ & + \Big(\beta_2+ \beta_1J+\beta_2K\Big)\Big( (y+1)I_{s,i}^{x,y+1,z-1} -y I_{s,i}^{x,y,z} \Big) 
        \\ & + \gamma \Big(-(i+y+2z)I_{s,i}^{x,y,z} + (i+1)I_{s-1,i+1}^{x,y,z} \\
        & \qquad\qquad + (y+1)I_{s,i}^{x,y+1,z-1} + 2(z+1)I_{s,i}^{x,y-1,z+1} \Big)
    \end{split}
    \label{eq:eff_degree}
\end{equation}

The coefficients $A, B, C, D, E, F, G, H, J, K$ provide the closure of the system by expressing conditional infection probabilities in terms of the state variables:
\begin{equation}
    \begin{split}
        &A =  \frac{[I\underline{S}S]}{[SS]}  = \frac{ \sum  si S_{s,i}^{x,y,z}}{\sum sS_{s,i}^{x,y,z}}\hspace{1cm}
        B =  \frac{[II\underline{S}S]}{[SS]} = \frac{ \sum  zs S_{s,i}^{x,y,z}}{\sum sS_{s,i}^{x,y,z}} \\
        & C =  \frac{[I\underline{S}SS]}{[SSS]} = \frac{ \sum  i x S_{s,i}^{x,y,z}}{\sum xS_{s,i}^{x,y,z}} \hspace{1cm}
        D =  \frac{[II\underline{S}SS]}{[SSS]} = \frac{ \sum  zx S_{s,i}^{x,y,z}}{\sum xS_{s,i}^{x,y,z}} \\
        & E =  \frac{[I\underline{S}IS]}{[ISS]}  = \frac{ \sum  iy S_{s,i}^{x,y,z}}{\sum yS_{s,i}^{x,y,z}} \hspace{1cm}
        F =  \frac{[II\underline{S}IS]}{[ISS]} = \frac{ \sum  zy S_{s,i}^{x,y,z}}{\sum yS_{s,i}^{x,y,z}} \\
        &G =  \frac{[I\underline{S}I]}{[IS]} = \frac{ \sum  i(i-1) S_{s,i}^{x,y,z}}{\sum iS_{s,i}^{x,y,z}} \hspace{1cm}
        H =  \frac{[II\underline{S}I]}{[IS]} = \frac{ \sum  zi S_{s,i}^{x,y,z}}{\sum iS_{s,i}^{x,y,z}}\\
        &J =  \frac{[I\underline{S}II]}{[IIS]} =\frac{ \sum  i z S_{s,i}^{x,y,z}}{\sum zS_{s,i}^{x,y,z}} \hspace{1cm} 
        K =  \frac{[II\underline{S}II]}{[IIS]}  =\frac{ \sum  z(z-1) S_{s,i}^{x,y,z}}{\sum zS_{s,i}^{x,y,z}}
    \end{split}
    \label{eq:closure_coefficients}
\end{equation}
Each coefficient represents the average infection pressure experienced by a particular type of neighborhood connection, weighted by the frequency of different configurations in the population.

The total counts of hyperedges in each state configuration can be expressed in terms of the node-level state variables, completing the closure of the system:
\begin{equation}
    \begin{split}
        & [SS] = \sum sS_{s,i}^{x,y,z}\\
        & [IS] = \sum iS_{s,i}^{x,y,z} = \sum sI_{s,i}^{x,y,z} \\
        & [II] = \sum iI_{s,i}^{x,y,z}\\
        & [SSS] = \sum xS_{s,i}^{x,y,z}\\
        & [ISS] = \sum yS_{s,i}^{x,y,z} = \sum xI_{s,i}^{x,y,z} \\
        & [IIS] = \sum zS_{s,i}^{x,y,z} = \sum yI_{s,i}^{x,y,z} \\
        & [III] = \sum zI_{s,i}^{x,y,z}
    \end{split}
    \label{eq:hyperedge_counts}
\end{equation}
where all summations are taken over all values of $s, i, x, y, z$ unless otherwise specified.

This completes the specification of our dynamical system, which can be numerically integrated to predict the temporal evolution of the epidemic on hypergraphs with both pairwise and higher-order interactions.

\subsection*{SIS Higher-order compact effective hyperdegree model, full derivation}\label{sec:SIS_compeffdegree_SM}

In principle, the effective degree model given in Equation~\eqref{eq:eff_degree} can be applied to describe epidemic processes on any configuration model network, including those with heterogeneous degree distributions, where most higher-order epidemic models typically fail. However, the complexity of the model grows substantially with degree heterogeneity, since the number of possible combinations of the variables $S_{s,i}^{x,y,z}$ and $I_{s,i}^{x,y,z}$ increases rapidly as the maximum degree grows. For networks with broad degree distributions, the full effective degree model can become computationally prohibitive.

An alternative approach to address this issue is to use the \textbf{compact effective degree model}, as presented in \cite{kiss2017mathematics}. This approximation significantly reduces the dimensionality of the state space while preserving the essential dynamics of the system.

In the compact formulation, we consider aggregate variables that group together all nodes with the same total degree at each interaction order, regardless of how many of those connections are to susceptible versus infected nodes. Specifically, we define:
\begin{itemize}
    \item \textbf{$S_{k_1}^{k_2}$}: the number of susceptible nodes with $k_1$ 1-hyperedges (pairwise connections) and $k_2$ 2-hyperedges (three-body interactions)
    \item \textbf{$I_{k_1}^{k_2}$}: the number of infected nodes with the same degree configuration
\end{itemize}
The total number of nodes with degree $(k_1, k_2)$ is denoted by $N_{k_1}^{k_2} = S_{k_1}^{k_2} + I_{k_1}^{k_2}$, which remains constant throughout the epidemic dynamics.
The connection between the detailed effective degree variables and the compact variables is established through a multinomial distribution approximation. We assume that the distribution of neighbors by state follows the global proportions in the network:
\begin{equation}
    S_{s,i}^{x,y,z} \approx \left[ \frac{k_1!}{s! \, i!} \langle I \rangle^i \langle S \rangle^s \right] \left[ \frac{k_2!}{x! \, y! \, z!} \langle X \rangle^x \langle Y \rangle^y \langle Z \rangle^z\right]  S_{k_1}^{k_2}
    \label{eq:multinomial_approx}
\end{equation}
where $k_1 = s + i$ and $k_2 = x + y + z$. This approximation implies that the number of hyperedges in each infection state follows a multinomial distribution, with probabilities determined by the global infection levels.

The probabilities $\langle S \rangle$, $\langle I \rangle$, $\langle X \rangle$, $\langle Y \rangle$, and $\langle Z \rangle$ represent the expected fractions of different hyperedge types and are determined by the global proportions of hyperedges in each state:
\begin{equation}
    \begin{split}
        &\langle I \rangle = \frac{[SI]}{[SS]+[SI]}, \qquad \langle S \rangle = \frac{[SS]}{[SS]+[SI]} \\[0.5em]
        &\langle X \rangle = \frac{[SSS]}{[SSS]+[SSI]+[SII]}, \quad \langle Y \rangle = \frac{[SSI]}{[SSS]+[SSI]+[SII]}, \quad \langle Z \rangle =\frac{[SII]}{[SSS]+[SSI]+[SII]}
    \end{split}
    \label{eq:global_probabilities}
\end{equation}
Note that these probabilities satisfy the normalization conditions: $\langle S \rangle + \langle I \rangle = 1$ and $\langle X \rangle + \langle Y \rangle + \langle Z \rangle = 1$.

To derive the evolution equations for the compact variables, we sum over all possible neighborhood configurations that correspond to the same total degree. The evolution of $S_{k_1}^{k_2}$ and $I_{k_1}^{k_2}$ is obtained by aggregating the infection and recovery events across all detailed states:
\begin{equation}
\begin{split}
    \dot{S}_{k_1}^{k_2} &= \sum_{s+i=k_1} \sum_{x+y+z=k_2} 
     \Big[ -(\beta_1 i + \beta_2 z) S_{s,i}^{x,y,z} + \gamma I_{s,i}^{x,y,z}\Big]\\[0.5em]
     \dot{I}_{k_1}^{k_2} &= \sum_{s+i=k_1} \sum_{x+y+z=k_2} 
     \Big[ (\beta_1 i + \beta_2 z) S_{s,i}^{x,y,z} - \gamma I_{s,i}^{x,y,z}\Big]
\end{split}
\label{eq:compact_evolution_sum}
\end{equation}
These equations capture only the direct infection and recovery events that change the infection state of nodes, while preserving their degree configuration. Crucially, terms corresponding to neighborhood rearrangements (which change $s, i, x, y, z$ but not $k_1, k_2$) cancel out in the summation.

In addition to tracking node states, the compact model requires explicit evolution equations for the hyperedge state variables. These equations describe how the distribution of infection states within hyperedges changes over time due to infection and recovery events.

For pairwise hyperedges, we track:
\begin{equation}
    \begin{split}
        \dot{[SI]} &= \gamma ([II]-[SI]) + \beta_1\Big( [I\underline{S}S] - [I\underline{S}I] - [SI]\Big) + \beta_2 \Big( [II\underline{S}S] - [II\underline{S}I] \Big) \\[0.5em]
        \dot{[II]} &= -2\gamma [II] + \beta_1\Big( [I\underline{S}I] +[SI]\Big) + \beta_2 \Big(  [II\underline{S}I] \Big)
    \end{split}
    \label{eq:pairwise_hyperedge_dynamics}
\end{equation}
For three-body hyperedges:
\begin{equation}
    \begin{split}
        \dot{[SSI]} &= \gamma ([SII] - [SSI]) + \beta_1\Big( [I\underline{S}SS] - [I\underline{S}SI]\Big) + \beta_2 \Big( [II\underline{S}SS] - [II\underline{S}SI] \Big) \\[0.5em]
        \dot{[SII]} &= \gamma ([III]-2[SII]) + \beta_1\Big( [I\underline{S}SI] - [I\underline{S}II]\Big) \\
        &\quad + \beta_2 \Big( [II\underline{S}SI] - [II\underline{S}II] - [SII] \Big) \\[0.5em]
        \dot{[III]} &= -3\gamma [III] + \beta_1\Big( [I\underline{S}II]\Big) + \beta_2 \Big( [II\underline{S}II] + [SII] \Big)
    \end{split}
    \label{eq:triadic_hyperedge_dynamics}
\end{equation}
These equations account for: (1) recovery events that change the infection state composition of hyperedges, and (2) infection events in neighboring hyperedges that indirectly affect the focal hyperedge through shared nodes (captured by the composite motif terms).

To close the system, we approximate the composite motif counts (such as $[I\underline{S}S]$, $[II\underline{S}SS]$, etc.) as functions of the global probabilities $\langle S \rangle$, $\langle I \rangle$, $\langle X \rangle$, $\langle Y \rangle$, and $\langle Z \rangle$, along with aggregate degree-based quantities. This mean-field approximation assumes that the probability of a node in a given state having particular types of neighboring connections is independent and determined by global proportions.

Applying the multinomial approximation and mean-field closure, the complete compact higher-order effective degree model reduces to:
\begin{equation}
    \begin{split}
    &\dot{S}_{k_1}^{k_2} = - \Big(\beta_1 k_1 \langle I \rangle + \beta_2 k_2 \langle Z \rangle\Big) S_{k_1}^{k_2} + \gamma \big( N_{k_1}^{k_2} - S_{k_1}^{k_2}\big)\\[0.5em]
    &\dot{[SI]} = \gamma[II]-(\gamma +\beta_1) [SI] + \Big( 1 - 2\langle I \rangle\Big) \Big(\beta_1\langle I \rangle D + \beta_2\langle Z \rangle C\Big)\\[0.5em]
    &\dot{[II]} = -2\gamma[II] + \beta_1[SI] + \langle I \rangle \Big(\beta_1\langle I \rangle D + \beta_2\langle Z \rangle C\Big)\\[0.5em]
    &\dot{[SSI]} = \gamma \Big( [SII]-[SSI] \Big) + \Big( 1 - \langle Z \rangle -2\langle Y \rangle\Big) \Big(\beta_1\langle I \rangle C + \beta_2\langle Z \rangle E\Big) \\[0.5em]
    &\dot{[SII]} = \gamma[III] -(2\gamma +\beta_2) [SII] + \Big( \langle Y \rangle - \langle Z \rangle\Big) \Big(\beta_1\langle I \rangle C + \beta_2\langle Z \rangle E\Big)\\[0.5em]
    & \dot{[III]} = -3\gamma[III] + \beta_2 [SII] + \langle Z \rangle \Big(\beta_1\langle I \rangle C + \beta_2\langle Z \rangle E\Big)
    \end{split}
    \label{eq:compact_final}
\end{equation}
with the probabilities defined as:
\begin{equation}
     \langle I \rangle = \frac{[SI]}{A}, \qquad  \langle Z \rangle = \frac{[SII]}{B}, \qquad  \langle Y \rangle = \frac{[SSI]}{B}
     \label{eq:probabilities_compact}
\end{equation}
and the aggregate quantities:
\begin{equation}
    \begin{split}
        &A = \sum_{k_1,k_2} k_1 S_{k_1}^{k_2}, \qquad B = \sum_{k_1,k_2} k_2 S_{k_1}^{k_2} \\[0.5em]
        &C = \sum_{k_1,k_2} k_1 k_2 S_{k_1}^{k_2}, \qquad D = \sum_{k_1,k_2} k_1 (k_1 - 1)S_{k_1}^{k_2}, \qquad E = \sum_{k_1,k_2} k_2 (k_2-1) S_{k_1}^{k_2}
    \end{split}
    \label{eq:aggregate_quantities}
\end{equation}

This compact formulation provides an efficient framework for analyzing epidemic dynamics on heterogeneous hypergraphs while maintaining the key features of higher-order transmission mechanisms.

\subsection*{Dimensionality Reduction of the Compact Effective Hyperdegree Model}
\label{sec:dimentionality_SM}

To quantify the computational advantage of the compact effective degree model, we provide a detailed analysis of the state-space reduction achieved by aggregating over detailed neighborhood configurations. This analysis demonstrates that the compact model offers substantial computational savings, particularly for networks with large maximum degrees.

Consider a hypergraph where nodes have a joint degree $(k_1,k_2)$ with bounded support:
\begin{equation}
0 \le k_1 \le K_1, \qquad 0 \le k_2 \le K_2
\end{equation}
where $K_1$ and $K_2$ represent the maximum degrees for pairwise and three-body interactions, respectively. We assume an arbitrary joint distribution $P(k_1,k_2)$ over this support.

In the full effective degree framework, we track compartments $S^{x,y,z}_{s,i}$ representing susceptible nodes with neighborhood vectors $[s,i,x,y,z]$. These variables are subject to the degree constraints:
\begin{equation}
s+i = k_1, \qquad x+y+z = k_2
\end{equation}
with all variables being non-negative integers. For a fixed joint degree $(k_1,k_2)$, the number of admissible states corresponds to the number of ways to partition $k_1$ into $(s,i)$ and $k_2$ into $(x,y,z)$. The number of ways to partition $k_1$ into two non-negative integers is $(k_1+1)$, while the number of ways to partition $k_2$ into three non-negative integers is $\binom{k_2+2}{2}$. Therefore, for each joint degree class, the number of states is:
\begin{equation}
(k_1+1)\binom{k_2+2}{2}.
\end{equation}
Summing over all possible joint degree classes gives the total state-space size of the full model:
\begin{align}
N_{\mathrm{full}}
&= \sum_{k_1=0}^{K_1}\sum_{k_2=0}^{K_2} (k_1+1)\binom{k_2+2}{2} \nonumber\\
&= \left(\sum_{k_1=0}^{K_1}(k_1+1)\right) \left(\sum_{k_2=0}^{K_2}\binom{k_2+2}{2}\right) \nonumber\\
&= \frac{(K_1+1)(K_1+2)}{2}\binom{K_2+3}{3},
\label{eq:full_state_space}
\end{align}
where we have used the identities $\sum_{k=0}^{K}(k+1) = \frac{(K+1)(K+2)}{2}$ and $\sum_{k=0}^{K}\binom{k+2}{2} = \binom{K+3}{3}$.

In the compact formulation, we track only the aggregate variables $S^{k_2}_{k_1}$, which represent the total number of susceptible nodes with joint degree $(k_1,k_2)$, regardless of their detailed neighborhood composition. There is exactly one state variable for each joint degree $(k_1,k_2)$, so the total number of states in the reduced model is simply:
\begin{equation}
N_{\mathrm{reduced}} = (K_1+1)(K_2+1).
\label{eq:reduced_state_space}
\end{equation}

The ratio of the state-space sizes quantifies the computational advantage of the compact model:
\begin{equation}
\frac{N_{\mathrm{full}}}{N_{\mathrm{reduced}}}
= \frac{(K_1+2)(K_2+2)(K_2+3)}{12}.
\label{eq:reduction_factor}
\end{equation}
This reduction factor grows rapidly with the maximum degrees. For example, with $K_1 = 20$ and $K_2 = 10$, the reduction factor is approximately 220, meaning the full model requires tracking 220 times more state variables than the compact model.

The asymptotic computational complexity of each model can be characterized by examining how the state-space size grows with the maximum degrees. The full effective-degree system scales as:
\begin{equation}
N_{\mathrm{full}} = \mathcal{O}(K_1^2 K_2^3),
\end{equation}
This cubic dependence on $K_2$ arises from the three-way partition $(x,y,z)$ and represents a significant computational burden for networks with high-degree nodes in the three-body interaction layer. In contrast, the compact model scales as:
\begin{equation}
N_{\mathrm{reduced}} = \mathcal{O}(K_1 K_2).
\end{equation}
This linear scaling in both maximum degrees makes the compact model dramatically more tractable for heterogeneous networks with broad degree distributions. The reduction from $\mathcal{O}(K_1^2 K_2^3)$ to $\mathcal{O}(K_1 K_2)$ represents a polynomial improvement in computational complexity, enabling analysis of epidemic dynamics on realistic hypergraphs that would be computationally prohibitive with the full model.

\end{document}